\documentclass[format=acmsmall, review=false, screen=true]{acmart}

\usepackage{booktabs} 

\usepackage{amssymb}

\usepackage{float}

\usepackage{graphicx}

\usepackage[skins]{tcolorbox}

\usepackage{latexsym}

\usepackage{amssymb}

\usepackage{graphicx}
\usepackage{multicol}
\usepackage{enumerate}
\usepackage{multirow}
\usepackage{subcaption}
\usepackage{setspace}

\usepackage[noend]{algpseudocode} 

\usepackage{algorithm}

\usepackage{algpseudocode}

\usepackage{colortbl}
\usepackage{color}
\usepackage{ctable}
\usepackage{wrapfig}

\usepackage{array}
\usepackage{microtype}
\DisableLigatures[f]{encoding = *, family = * }

\usepackage{helvet}         
\usepackage{courier}        
\usepackage{type1cm}        
\usepackage{graphicx}        
\usepackage{epsfig}
\usepackage{multicol}        
\usepackage[bottom]{footmisc}
\usepackage{alltt}

\usepackage{tabularx,multicol,multirow}

\usepackage{caption}

\usepackage{verbatim}

\acmJournal{arXiv}

\acmDOI{0000001.0000001}

\begin{document}
\title{Modelling and Analysis of Temporal Preference Drifts Using A Component-Based Factorised Latent Approach} 

\author{F. Zafari}
\orcid{1234-5678-9012-3456}
\affiliation{%
	\institution{Swinburne University of Technology}
	\streetaddress{John St, Hawthorn}
	\city{Melbourne}
	\state{VIC}
	\postcode{3122}
	\country{Australia}}
\email{fzafari@swin.edu.au}
\author{I. Moser}
\affiliation{%
	\institution{Swinburne University of Technology}
	\streetaddress{John St, Hawthorn}
	\city{Melbourne}
	\state{VIC}
	\postcode{3122}
	\country{Australia}}
\email{imoser@swin.edu.au}
\author{T. Baarslag}
\affiliation{%
	\institution{Centrum Wiskunde \& Informatica}
	\streetaddress{}
	\city{Amsterdam}
	\state{}
	\postcode{}
	\country{Netherlands}}
\email{T.Baarslag@cwi.nl}

\begin{abstract}
In recommender systems, human preferences are identified by a number of individual components with complicated interactions and properties. Recently, the dynamicity of preferences has been the focus of several studies. The changes in user preferences can originate from substantial reasons, like personality shift, or transient and circumstantial ones, like seasonal changes in item popularities. Disregarding these temporal drifts in modelling user preferences can result in unhelpful recommendations. Moreover, different temporal patterns can be associated with various preference domains, and preference components and their combinations. These components comprise preferences over features, preferences over feature values, conditional dependencies between features, socially-influenced preferences, and bias. For example, in the movies domain, the user can change his rating behaviour (bias shift), her preference for genre over language (feature preference shift), or start favouring drama over comedy (feature value preference shift). In this paper, we first propose a novel latent factor model to capture the domain-dependent component-specific temporal patterns in preferences. The component-based approach followed in modelling the aspects of preferences and their temporal effects enables us to arbitrarily switch components on and off. We evaluate the proposed method on three popular recommendation datasets and show that it significantly outperforms the most accurate state-of-the-art static models. The experiments also demonstrate the greater robustness and stability of the proposed dynamic model in comparison with the most successful models to date. We also analyse the temporal behaviour of different preference components and their combinations and show that the dynamic behaviour of preference components is highly dependent on the preference dataset and domain. Therefore, the results also highlight the importance of modelling temporal effects but also underline the advantages of a component-based architecture that is better suited to capture domain-specific balances in the contributions of the aspects.
\end{abstract}


\maketitle

\renewcommand\shortauthors{Zafari et al.}

\section{Introduction}
\label{Introduction}

Recommender systems suggest items (movies, books, music, news, services, etc.) that appear most likely to interest a particular user. Matching users with the most desirable items helps enhance user satisfaction and loyalty. Therefore, many e-commerce leaders such as Amazon and Netflix have made recommender systems a salient part of their services \citep{koren2009matrix}.
Currently, most recommendation techniques leverage user-provided feedback data to infer user preferences \citep{chen2015recommender}.
Typically, recommender systems are based on collaborative filtering (CF) \citep{koren2011advances, aldrich2011recommender}, where the preferences of a user are predicted by collecting rating information from other similar users or items \citep{ma2008sorec}.
Many recent studies have contributed extensions to the basic Probabilistic Matrix Factorisation (PMF) by incorporating additional information.
Despite their popularity and good accuracy, recommender systems based on latent factor models encounter some important problems in practical applications \citep{zafari2016feature}. In these models, it is assumed that all values for item features are equally preferred by all users.

Another major problem with latent factor models based on matrix factorisation is that they do not usually take conditional preferences into consideration \citep{liu2015conditional}.
Furthermore, in general, latent factor models do not consider the effect of social relationships on user preferences, which encompasses peer selection (homophily) and social influence \citep{Lewis03012012, zafarani2014social}. 
In previous work, we addressed the problem of modelling the socially-influenced conditional feature value preferences, and proposed CondTrustFVSVD \citep{zafari2017modelling}.

Since data usually changes over time, the models should continuously update to reflect the present state of data \citep{koren2010collaborative}.
A major problem with the most of the recent recommender systems is that they mostly ignore the drifting nature of preferences \citep{zafari2017modelling}. Modelling the time drifting data is a central problem in data mining. Drifting preferences can be considered a particular type of concept drift, which has received much attention from researchers in recent years \citep{widmer1996learning}. However, very few recommendation models have considered the drifting nature of preferences \citep{chatzis2014dynamic}. Changes in user preferences can originate from substantial reasons, or transient and circumstantial ones. For example, the items can undergo \textit{seasonal changes} or some items may experience \textit{periodic changes}, for instance, become popular in the specific holidays.

Apart from the short-term changes, user preferences are also subject to long term drifts. For example, a user may be a fan of romantic or action movies at a younger age, while his/her preference may shift more towards drama movies as gets older. Also, users may change their rating scale over time. For example, a user may be very strict and give 3 out of 5 for the best movie. However, might become less strict with age and be more willing to elect the full rate when fully satisfied. A similar situation may apply for movies. A movie may receive a generally high/low rate at some time period, and lower/higher rates at some other period \citep{koren2010collaborative}. Therefore, a preference model should be able to distinguish between different types of preference drifting, and model them individually in order to achieve the highest accuracy.

In recommender systems research, six major aspects to the preferences have been identified. These aspects include \textit{feature preferences} (\citep{zafari2015dopponent,salakhutdinov2011probabilistic}), \textit{feature value preferences} (\citep{zafari2016popponent,zafari2017popponent,zhang2014explicit}), \textit{socially-influenced preferences} (\citep{zafari2017modelling,zhao2015probabilistic,ma2008sorec,ma2011recommender,jamali2010matrix}), \textit{temporal dynamics} (\citep{koren2010collaborative}), \textit{conditional preferences} (\citep{liu2015conditional}), and \textit{user and item biases} (\citep{koren2011advances}). Feature value preferences refer to the relative favourability of each one of the item feature values, social influence describes the influence of social relationships on the preferences of a user, temporal dynamics means the drift of the preferences over time, conditional preferences refer to the dependencies between item features and their values, and user and item biases pertain to the systematic tendencies for some users to give higher ratings than others, and for some items to receive higher ratings than others (\citep{koren2011advances}). Modelling the temporal properties of these preference aspects is the central theme of this paper.

In this paper, we extend our previous work \citep{zafari2017modelling}, by considering the drifting nature of preferences and their constituting aspects. We assume that the socially-influenced preferences over features and conditional preferences over feature values, as well as user and item rating scales can be subject to temporal drift.
Therefore, the two major research questions addressed in this paper are:

\begin{itemize}
	\item How can we efficiently model the drifting behaviour 
	preferences, and how much improvement would incorporating such information make?
	\item Which aspects are more subject temporal changes, and how is this related to the domain on which the model is trained?
\end{itemize}

The current work proposes a novel latent factor model based on matrix factorisation to address these two questions.
This paper has two major contributions for the field. 
In this paper, we make further improvements on the accuracy of, CondTrustFVSVD, a model that we proposed earlier. CondTrustFVSVD proved to be the the most accurate model among a large set of state of the art models. The additional improvements were achieved by incorporating the \textit{temporal dynamics of preference aspects}. We also draw conclusions about the dynamicity of preference aspects, by analysing the temporal aspects of the these aspects using a component-based approach, and show which aspects are more subject to drift over time. This research provides useful insights into the \textit{accurate modelling of preferences and their temporal properties}, and helps pave the way for boosting the performance of recommender systems. The findings suggest that the temporal aspects of user preferences can vary from one domain to another. Therefore, \textit{modelling domain-dependent temporal effects of preference aspects} are critical in improving the quality of recommendations.

The rest of the paper is organised as follows: The related work is introduced in section \ref{Related Work}. In section \ref{Brief introduction of PMF and CondTrustFVSVD}, we first briefly introduce probabilistic matrix factorisation, and CondTrustFVSVD. Then in section \ref{Time-Aware CondTrustFVSVD (Aspect-MF)} we introduce Aspect-MF to overcome the challenge of learning drifting conditional socially-influenced preferences over feature values. In section \ref{Experiments}, we first explain the experimental setup, and then report on the results of Aspect-MF using two popular recommendation datasets. Finally we conclude the paper in section \ref{Conclusion and Future Work}, by summarising the main findings and giving the future directions of this work.

\section{Related work}
\label{Related Work}
Collaborative Filtering models are broadly classified into memory-based and model-based approaches. Memory- or instance-based learning methods predict the user preferences based on the preferences of other users or the similarity of the items. Item-based approaches in memory-based CF \citet{d2015sentiment} calculate the similarity between the items, and recommend the items similar to the items that the user has liked in the past. User-based approaches recommend items that have been liked by similar users \citet{ma2008sorec}.
The time-dependent collaborative filtering models are also classified into the memory-based time-aware recommenders and model-based time-aware recommenders \citep{xiang2009time}.

\subsection{Model-based time-aware recommenders}
The models in this category usually fall into four classes: 1) models based on Probabilistic Matrix Factorisation, 2) models based on Bayesian Probabilistic Matrix Factorisation, and 3) models based on Bayesian Probabilistic Tensor Factorisation, and 4) models based on Bayesian Probabilistic Tensor Factorisation.

\subsubsection{Models based on probabilistic matrix factorisation}
Modelling the drifting preferences using a model-based approach based on PMF has first been considered by Koren \citep{koren2010collaborative} in TimeSVD++. TimeSVD++ builds on the previous model called SVD++ \citep{koren2009matrix}, in which the user preferences are modelled through a latent factor model that incorporates the user bias, item bias, and also the implicit feedback given by the users. 
For each one of these preference aspects, Koren \citep{koren2010collaborative} used a time-dependent factor to capture both transient and long-term shifts. They showed TrustSVD++ achieves significant improvements over SVD++ on a daily granularity \citep{xiang2009time}.

In TrustFVSVD \citep{zafari2017modelling}, we extended TrustSVD by adding the preferences over feature values and the conditional dependencies between the features. We did this by adding additional matrices that captured the feature value discrepancies, where the values of these matrices were related to the values of the social influence matrix. In TrustFVSVD, the explicit influence of the social relationships on each one of the aspects of preferences were captured. Through comprehensive experiments on three benchmark datasets, we showed that TrustFVSVD significantly outperformed TrustSVD and a large set of state of the art models. However, similar to most of the state of the art models, in TrustFVSVD, we assumed that the preferences are static.

Another model-based time-aware recommendation model was proposed by Koenigstein, Dror and Koren \citep{koenigstein2011yahoo}. In this model, the authors use session factors to model specific user behaviour in music learning sessions.
Unlike TimeSVD++ which is domain-independent, was developed especially for the music domain. First, it enhances the bias values in SVD++, by letting the item biases share components for items linked by the taxonomy. For example, the tracks in a good album may all be rated higher than the average, or a popular artist may receive higher ratings than the the average for items. Therefore, shared bias parameters are added to different items with a common ancestor in the taxonomy hierarchy of the items. Similarly, the users may also tend to rate artists or genres higher than songs. Therefore, the user bias is also enhanced by adding the type of the items. It is also assumed that unlike in the movies domain, in music it is common for the users to listen to many songs, and rate them consecutively. Such ratings might be rated similarly due to many psychological phenomena.
The advantage of the models proposed by Koenigstein, Dror and Koren \citep{koenigstein2011yahoo} and Koren \citep{koren2010collaborative} that extend SVD++ is that they enable the capturing of dynamicity of the preference aspects with a high granularity for aspects that are assumed to be more subject to temporal drift. Furthermore, as shown in \citep{koenigstein2011yahoo}, domain-dependent temporal aspects of the preferences and their individual aspects can also be taken into consideration.

Jahrer, Toscher and Legenstein \citep{jahrer2010combining} split the rating matrix into several matrices, called bins, based on their time stamps. For each bin, a separate time-unaware model is trained by producing an estimated rating value that is obtained using the ratings of given for that bin. Each one of the bins are assigned a weight value, and the final rating is obtained by combining the ratings that are obtained through the models trained on each bin. Therefore, using this approach, they combine multiple time-unaware models into a single time-aware model. The disadvantage of this model is that the ratings matrix is usually sparse as it is, and it even becomes sparser, when the ratings are split into bins.

A similar approach is followed in the model proposed by Liu and Aberer \citep{liu2013soco}. They systematically integrated contextual information and social network information into a matrix factorization model to improve the recommendations. To overcome the sparsity problem of training separate models based on their time-stamps, they applied a random decision trees algorithm, and create a hierarchy of the time-stamps. For example, the ratings can be split based on year in the first level, month in the second level, day in the third level, and so on. They argue that the ratings that are given at similar time intervals are better correlated with each other, and therefore such clustering is justified. They also added the influence of the social friends to the model, using a context-aware similarity function. In this function users who give similar ratings to those of their friends in similar contexts get higher similarity values. Consequently, in this model, the role of time on the social influence is also indirectly taken into consideration.

Baltrunas, Ludwig and Ricci \citep{baltrunas2011matrix} argued that methods based on tensor factorisation can improve the accuracy when the datasets are large. Tensor factorisation requires the addition of a large number of model parameters that must be learned. 
When the datasets are small, simpler models with fewer parameters can perform equally well or better. In their  method, a matrix is added to capture the influence of contextual factors (e.g. time) on the user preferences by modelling the interaction of contextual conditions with the items. 
Although the model is quite simple and fast, it does not include the effect of time on individual preference aspect. Unlike the models proposed by Koenigstein, Dror and Koren \citep{koenigstein2011yahoo} and Koren \citep{koren2010collaborative}, it can not capture fine-grained and domain-specific dynamicities.

\subsubsection{Models based on Bayesian probabilistic matrix factorisation}
BPMF extends the basic matrix factorisation \citep{salakhutdinov2008bayesian} by assuming Gaussian-Wishart priors on the user and item regularisation parameters and letting the hyper-parameters be trained along with the model parameters. Dynamic BPMF (dBPMF) is a non-parametric Bayesian dynamic relational data modelling approach based on the Bayesian probabilistic matrix \citep{luo2016bayesian}. This model imposes a dynamic hierarchical Dirichlet process (dHDP) prior over the space of probabilistic matrix factorisation models to capture the time-evolving statistical properties of modelled sequential relational datasets. The dHDP was developed to model the time-evolving statistical properties of sequential datasets, by linking the statistical properties of data collected at consecutive time points via a random parameter that controls their probabilistic similarity.

\subsubsection{Models based on probabilistic tensor factorisation}
In tensor factorisation methods, the context variables are modelled in the same way as the users and items are modelled in matrix factorisation techniques, by considering the interaction between users-items-context. In tensor factorisation methods, the three dimensional user-item-context ratings are factorised into three matrices, a user-specific matrix, an item-specific matrix, and a context-specific matrix.
A model in this category is proposed by Karatzoglou et al. \citep{karatzoglou2010multiverse}, who used Tensor Factorisation with CP-decomposition, and proposed multi-verse recommendation, which combines the data pertaining to different contexts into a unified model. Therefore, similar to the model proposed by Baltrunas, Ludwig and Ricci \citep{baltrunas2011matrix}, other contextual information besides time (e.g. user mode, companionship) can also be taken into consideration. However, unlike Baltrunas, Ludwig and Ricci \citep{baltrunas2011matrix}, they factorise the rating tensor into four matrices, a user-specific matrix, an item-specific matrix, a context-specific matrix, and a central tensor, which captures the interactions between each user, item, and context value. Then the original ratings tensor, which includes the ratings given by users to items in different contexts (e.g. different times) can be reconstructed by combining the four matrices back into the ratings tensor.
Other models in this category are the models proposed by Li et al. \citep{li2011tracking} and Pan et al. \citep{pan2013robust}.

\subsubsection{Models based on Bayesian probabilistic tensor factorisation}
There is a class of dynamic models that are based on Bayesian Probabilistic Tensor Factorisation (BPTF) \citep{xiong2010temporal}. BPTF generalises BPMF by adding tensors to the matrix factorisation process. A tensor extends the two dimensions of the matrix factorisation model to three or more dimensions. Therefore, besides capturing the user-specific and item-specific latent matrices, this model also trains a time-specific latent matrix, which captures the latent feature values in different time periods. The models based on tensor factorisation are similar in introduction of the time-specific matrices into the factorisation process. However, they are different in the way they factorise the ratings matrix into the user, item, and time matrices, and also the way they train the factorised matrices. Similar to BPMF, BPTF uses Markov Chain Monte Carlo with Gibbs sampling to train the factorised matrices. 

\subsection{Memory-based time-aware recommenders}
Some simple time-dependent collaborative filtering models have been proposed by Lee, Park and Park \citep{lee2008time}. The models use item-based and user-based collaborative filtering, and exploit a pseudo-rating matrix, instead of the real rating matrix. In the pseudo-rating matrix the entries are obtained using a rating function, which is defined as the rating value when an item with launch time $l_j$ was purchased at time $p_i$. This function was inspired by two observations, that more recent purchases better reflected a user's current preferences, and also recently launched items appealed more to the users. If the users are more sensitive to the item's launch time, the function gives more weight to new items, and if the user's purchase time is more important in estimating their current preference, the function assigns more weight to recent purchases. After obtaining the pseudo-rating matrix, the neighbours are obtained as in the traditional item-based or user-based approaches, and the items are recommended to the users. These models are less related to the proposed model in this paper, so we are not going to review them further.

\section{Modelling time-aware preference aspects in CondTrustFVSVD}
\label{Modelling time-aware preference aspects in CondTrustFVSVD}

In this section, we explain how to integrate the time-awareness on different aspects of preferences into CondTrustFVSVD \citep{zafari2017modelling}.

\subsection{Brief introduction of PMF and CondTrustFVSVD}
\label{Brief introduction of PMF and CondTrustFVSVD}

In rating-based recommender systems, the observed ratings are represented by the user-item ratings matrix $R$, in which the element $R_{uj}$ is the rating given by the user $u$ to the item $j$. Usually, $R_{uj}$ is a 5-point integer, 1 point means very bad, and 5 points means excellent. Let $P \in \mathbb{R}^{N \times D}$ and $Q \in \mathbb{R}^{M \times D}$ be latent user and item feature matrices, with vectors $P_{u}$ and $Q_{j}$ representing user-specific and item-specific latent feature vectors respectively ($N$ is the number of users, $M$ is the number of items, and $D$ is the number of item features). In PMF, $R_{uj}$ is estimated by the inner product of the latent user feature vector $P_u$ and latent item feature vector $Q_j$, that is $\hat{R}_{uj} = P_uQ_j^T$.

PMF maximises the log-posterior over the user and item latent feature matrices with rating matrix and fixed parameters given by Eq. \ref{eq1}.

\small
\begin{equation}
	\begin{split}
		\label{eq1}
		\ln p( \, P,Q|R,\sigma,\sigma_{P},\sigma_{Q}) \, 
		= 
		\ln p( \, R|P,Q,\sigma) \, 
		+ 
		\ln p( \, P|\sigma_{P}) \,
		+
		\ln p( \, Q|\sigma_{Q}) \,
		+
		C
	\end{split}
\end{equation}
\normalsize

where $C$ is a constant that is not dependent on $P$ and $Q$. $\sigma_{P}$, $\sigma_{Q}$, and $\sigma$ are standard deviations of matrix entries in $P$, $Q$, and $R$ respectively. Maximising the log-posterior probability in Eq. \ref{eq1} is equivalent to minimising the error function in Eq. \ref{eq2}.

\small
\begin{equation}
	\begin{split}
		\label{eq2}
		argmin_{U,V}
		[ \,
		E 
		= \frac{1}{2} \sum_{u=1}^{N}\sum_{j=1}^{M}I_{uj} (\, R_{uj} - \hat{R}_{uj} )\,^2
		+
		\frac{\lambda_{P}}{2} \sum_{u=1}^{N}\|P_{u}\|_{Frob}^{2}
		+
		\frac{\lambda_{Q}}{2} \sum_{j=1}^{M}\|Q_{j}\|_{Frob}^{2}
		] \,
	\end{split}
\end{equation}
\normalsize

where $\|.\|_{Frob}$ denotes the Frobenius norm, and $\lambda_{P} = \frac{\sigma^2}{\sigma_{P}^2}$ and $\lambda_{Q} = \frac{\sigma^2}{\sigma_{Q}^2}$  (regularisation parameters). $Stochastic$ $Gradient$ $Descent$ and $Alternating$ $Least$ $Squares$ are usually employed to solve the optimisation problem in Eq. \ref{eq2}. Using these methods, the accuracy of the method measured on the training set is improved iteratively.

As mentioned in the introduction section, the disadvantage of traditional matrix factorisation methods is that the discrepancies between users in preferring item feature values and conditional dependencies between features are disregarded.
CondTrustFVSVD \citep{zafari2017modelling} addresses these problems by adding matrices $W$ and $Z$ to learn the preferences over item feature values.
Suppose that a social network is represented by a graph $\mathbb{G} = (\mathbb{V},\mathbb{E})$, where $\mathbb{V}$ includes a set of 
users (nodes) and $\mathbb{E}$ represents the trust relationships among the users (edges). We denote the adjacency matrix by $T \in \mathbb{R}^{N \times N}$, where $T_{uv}$ shows the degree to which user $u$ trusts user $v$. Throughout this paper, we use the indices $u$ and $v$ for the users and indices $i$ and $j$ for items, and indices $f$ and $f^{'}$ for item features.
In CondTrustFVSVD, all aspects of preferences are assumed to be subject to change by social interactions, and therefore the explicit influence of social relationships on each of the aspects of the preferences are modelled. 
In this method, we assume that the user preferences over an item feature can be formulated with a linear function. In this function, matrix $W$ is used to capture the "gradient" values and matrix $Z$ is used to learn the "intercept" values. 
These matrices have the same dimensions as the user matrix $P$.
According to this figure, the probabilities of the matrices $P$, $Q$, $W$, $Z$, $\omega$, $y$ and vectors $bu$ and $bi$ are dependent on the hyper-parameters $\sigma_{P}$, $\sigma_{Q}$, $\sigma_{W}$, $\sigma_{Z}$, $\sigma_{\omega}$, $\sigma_{y}$, $\sigma_{bu}$ and $\sigma_{bi}$ respectively. Likewise, the probability of obtaining the ratings in matrix $R$ is conditional upon the matrices $P$, $Q$, $W$, $Z$, $\omega$, $y$ and vectors $bu$ and $bi$. CondTrustFVSVD finds the solution for the optimisation problem formulated by Eq. \ref{eq3}.

\small
\begin{equation}
	\begin{split}
		\label{eq3}
		argmin_{P,Q,W,Z,\omega,y,bu,bi}
		[ \,
		E &
		=
		\frac{\lambda_t}{2} \sum_{u=1}^{N}\sum_{\forall v \in T_u}I_{uv} (\, T_{uv} - \sum_{f=1}^{D}P_{uf}\omega_{vf} )\,^2 
		+
		\frac{\lambda_t}{2} \sum_{u=1}^{N}\sum_{\forall v \in T_u}(\, T_{uv} - \sum_{f=1}^{D}(1 - W_{uf})\omega_{vf} )\,^2 \\ &
		+
		\frac{\lambda_t}{2} \sum_{u=1}^{N}\sum_{\forall v \in T_u}(\, T_{uv} - \sum_{f=1}^{D}Z_{uf}\omega_{vf} )\,^2 
		+
		\frac{1}{2} \sum_{u=1}^{N}\sum_{j=1}^{M} (\, R_{uj} - \hat{R}_{uj} )\,^2 \\ &
		+
		\sum_{u=1}^{N}(\frac{\lambda_P}{2}|I_u|^{-\frac{1}{2}} + \frac{\lambda_{T}}{2}|T_u|^{-\frac{1}{2}})\|P_{u}\|_{Frob}^{2}
		+
		\frac{\lambda_{Q}}{2} \sum_{j=1}^{M}\|Q_{j}\|_{Frob}^{2} \\ &
		+
		\sum_{u=1}^{N}(\frac{\lambda_W}{2}|I_u|^{-\frac{1}{2}} + \frac{\lambda_{T}}{2}|T_u|^{-\frac{1}{2}})\|W_{u}\|_{Frob}^{2}
		+
		\sum_{u=1}^{N}(\frac{\lambda_Z}{2}|I_u|^{-\frac{1}{2}} + \frac{\lambda_{T}}{2}|T_u|^{-\frac{1}{2}})\|Z_{u}\|_{Frob}^{2} \\ &
		+
		\frac{\lambda}{2} \sum_{i=1}^{M}|U_{i}|^{-\frac{1}{2}}\|y_i\|_{Frob}^{2}
		+
		\frac{\lambda_{\omega}}{2} \sum_{v=1}^{N}|T^{+}_{v}|^{-\frac{1}{2}}\|\omega_v\|_{Frob}^{2} \\ &
		+
		\frac{\lambda_{bu}}{2} \sum_{u=1}^{N}|I_u|^{-\frac{1}{2}}bu_{u}^{2}
		+
		\frac{\lambda_{bi}}{2} \sum_{j=1}^{M}|U_j|^{-\frac{1}{2}}bi_{j}^{2}
		+
		\frac{\lambda_{Y}}{2}\sum_{f=1}^{D}\sum_{f^{'}=1}^{D}Y_{ff^{'}}^2 
		] \,
	\end{split}
\end{equation}
\normalsize

where $\lambda_{W} = \frac{\sigma^2}{\sigma_{W}^2}$, $\lambda_{Z} = \frac{\sigma^2}{\sigma_{Z}^2}$, $\lambda_{\omega} = \frac{\sigma^2}{\sigma_{\omega}^2}$, $\lambda_{y} = \frac{\sigma^2}{\sigma_{y}^2}$, $\lambda_{bu} = \frac{\sigma^2}{\sigma_{bu}^2}$, $\lambda_{bi} = \frac{\sigma^2}{\sigma_{bi}^2}$, $\lambda_{Y} = \frac{\sigma^2}{\sigma_{Y}^2}$. $\mu$ denotes the global average of the observed ratings, and $bu_i$ and $bi_j$ denote biases for user $i$ and item $j$ respectively. $I_u$ is the set of items rated by user $u$ and $U_j$ is the set of users who have rated item $j$. The values of $\hat{R}_{uj}$ in Eq. \ref{eq3} are obtained using Eq. \ref{eq4}.

\begin{equation}
	\begin{split}
		\label{eq4}
		\hat{R}_{uj} = \mu + bu_u + bi_j + \sum_{f=1}^{D}(P_{uf} + |I_u|^{-\frac{1}{2}}\sum_{\forall i \in I_u}y_{if} + |T_u|^{-\frac{1}{2}}\sum_{\forall v \in T_u}\omega_{vf})(W_{uf}Q_{jf} + Z_{uf})
	\end{split}
\end{equation}

According to the Eq. \ref{eq4}, the user $u$'s preference value over an item $j$ is defined using different aspects. These aspects are user bias, item bias, the \textbf{socially-influenced preferences over features}, and the \textbf{socially-influenced preferences over feature values}. Therefore, preferences are defined using different aspects that interact with each other by influencing the values of one another.

\subsection{Time-aware CondTrustFVSVD (Aspect-MF)}
\label{Time-Aware CondTrustFVSVD (Aspect-MF)}
In the following sections, we first provide a high-level view of Aspect-MF by explaining the interactions between aspects that are captured by the model, and then elaborating how the aspects are trained from the users' ratings and social relationships.

\subsubsection{Aspect interactions and high-level view of the model}
\label{Aspect interactions and high-level view of the model}
To address the problem of capturing drifting socially-influenced conditional preferences over feature values, we extend the method CondTrustFVSVD, by adding the dynamicity of each one of the preference aspects that are assumed to be subject to concept drift. 
The method proposed here is abbreviated to Aspect-MF. A high-level overview of the preference aspects in Aspect-MF are presented in Fig. \ref{fig1}.

\begin{figure}[!ht]
	\setcounter{figure}{0}
	\vskip 0cm
	\centerline{\includegraphics[scale=0.7]{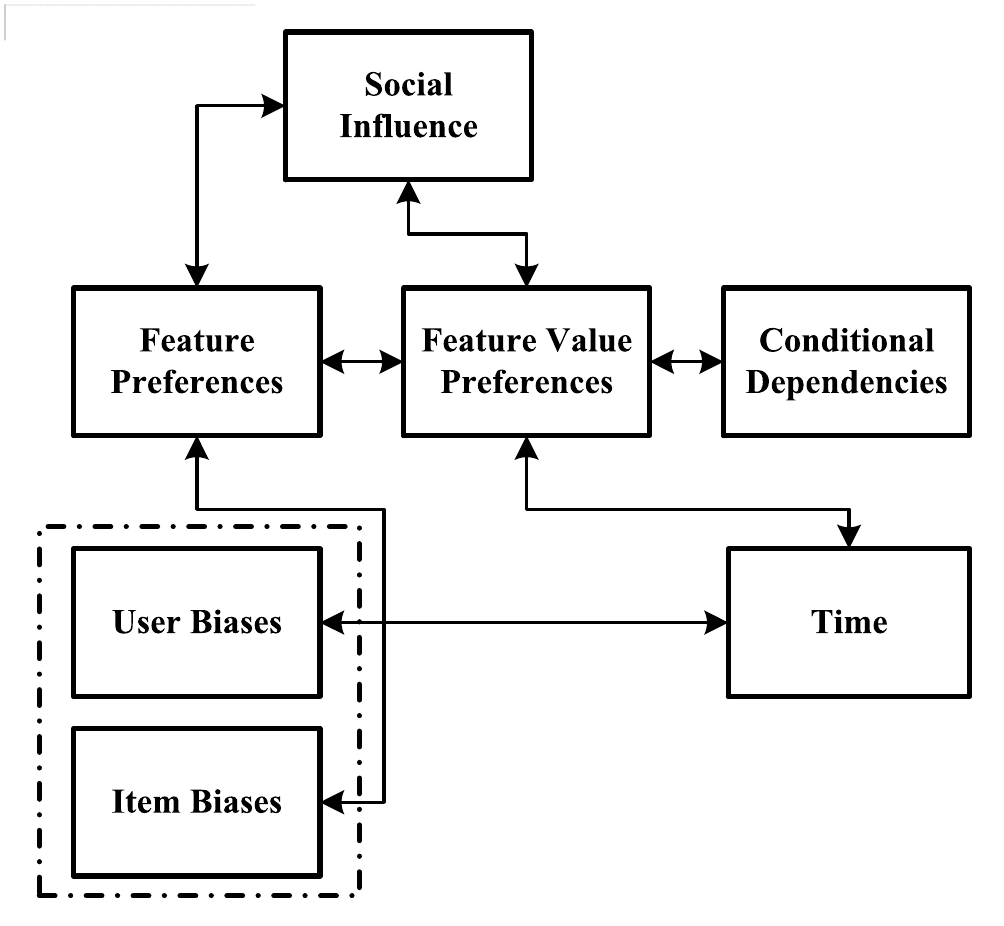}}
	\caption{The preference aspects and their interplay in Aspect-MF}
	\label{fig1}
\end{figure}

\begin{figure}[!ht]
	\vskip 0cm
	\centering
	\setcounter{figure}{1}
	\begin{subfigure}[b]{0.4\textwidth}
		\includegraphics[width=1\linewidth]{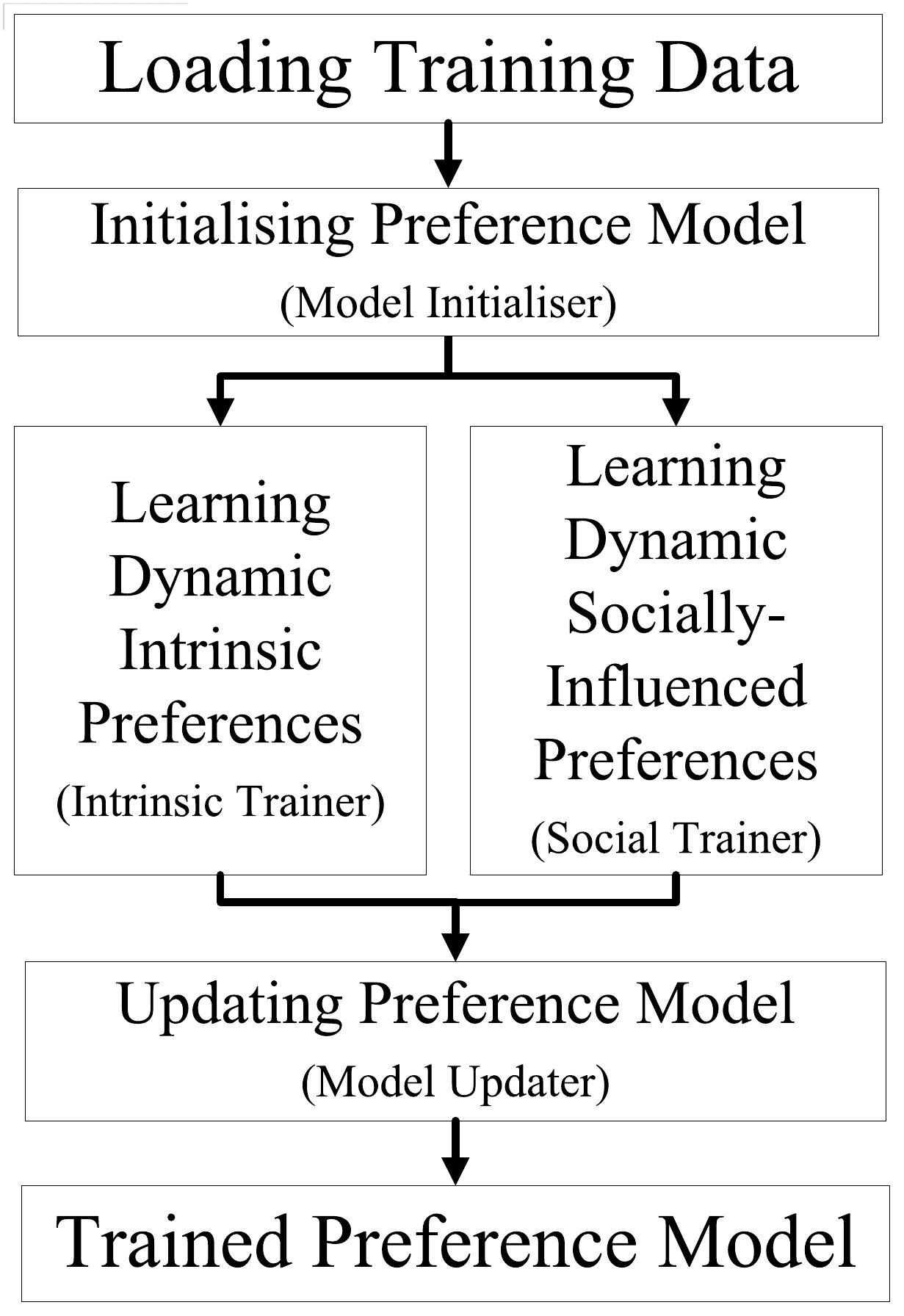}
		\caption{}
		\label{fig2a}
	\end{subfigure}%
	\begin{subfigure}[b]{0.6\textwidth}
		\includegraphics[width=1\linewidth]{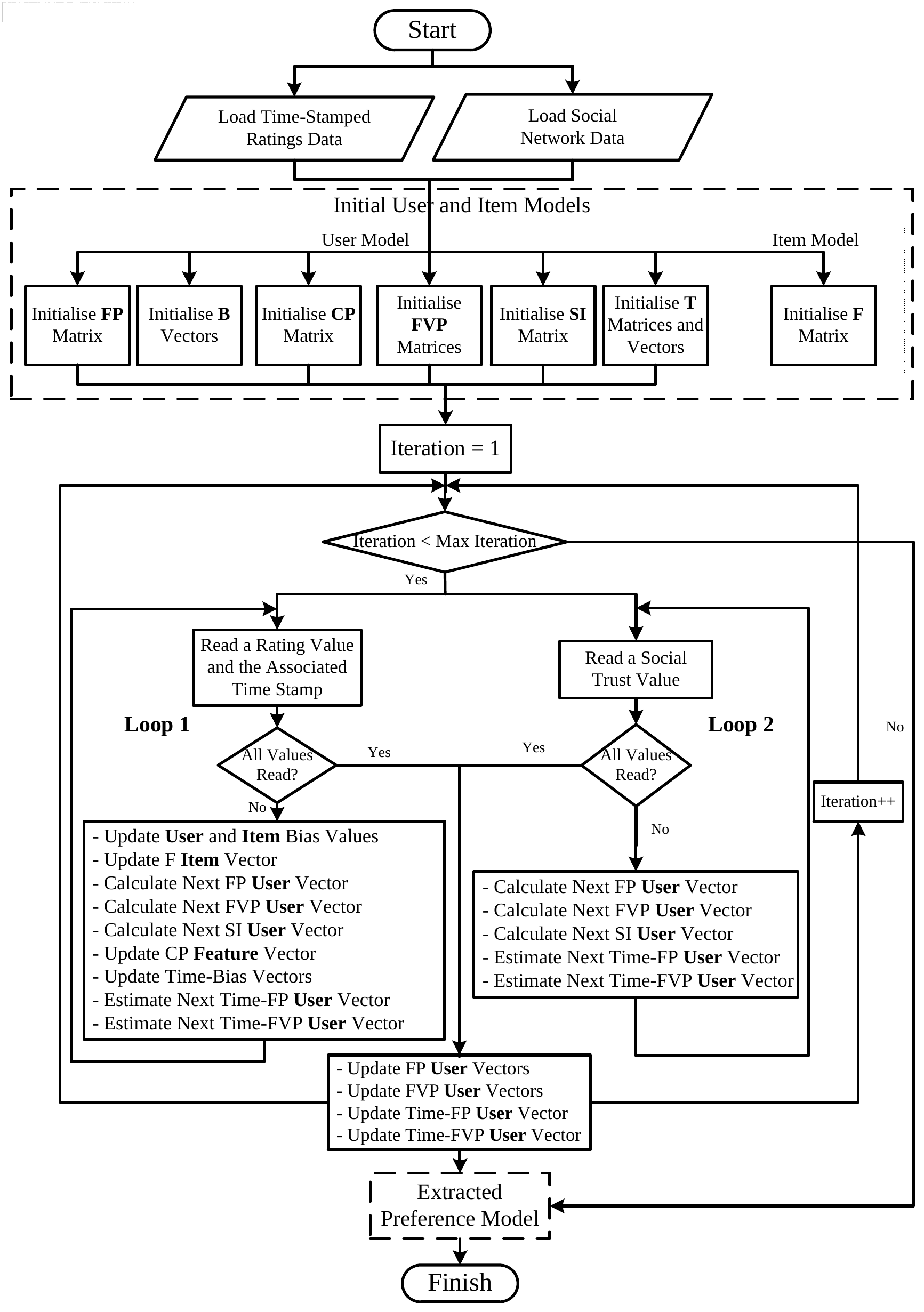}
		\caption{}
		\label{fig2b}
	\end{subfigure}%
	\centering
	\caption{a) The high-level representation of Aspect-MF and b) its flow chart}
	\label{fig2}
\end{figure}

In Fig. \ref{fig2a}, \textbf{FP} represents preferences over features, which is captured by matrix $P$ in the basic matrix factorisation. \textbf{F} represents item features captured by matrix $Q$ in the basic matrix factorisation. \textbf{CP} represents conditional dependencies, \textbf{FVP} represents preferences over feature values, \textbf{SI} stands for social influence, and finally \textbf{T} is an abbreviation for time.
Aspect-MF incorporates additional matrices and vectors into matrix factorisation to capture as many aspects present in the data as possible. As Fig. \ref{fig2} shows, the model starts by loading the time-stamped user ratings as well as the social network data into the memory.
The main loop accounts for the learning iterations over the model. The first loop within the main loop iterates over the time-stamped user-item ratings matrix, while the second loop iterates over the social network adjacency matrix, to train the socially influenced parts of the model. In each loop, one entry of the input matrix is read and used to update the matrices/vectors related to that input data. As can be seen, the user and item bias values are only updated in loop 1, since they are only related to the user-item ratings. Both user-item ratings and users' social relationships include information about the users' preferences over features. Therefore, the new values for FP are calculated in both loops and updated in the main loop, when all new values have been calculated. Similarly, the values for SI and FVP depend on both user-item ratings and social relationships. Consequently, their new values are calculated inside both loops 1 and 2, and are updated in the main loop. In contrast, the values of F as well as CP only need the user-item ratings to be updated. Therefore, they are immediately updated inside loop 1. The time aspect includes parameters that account for the dynamics of user and item biases, feature value preferences, and preferences over features. Since bias values do not depend on the user-item ratings matrix, they are updated immediately in loop 1. However, the new values for the dynamics of feature value preferences, and preferences over features are updated in the main loop. In Aspect-MF, every one of the preference aspects can be arbitrarily switched off and on by setting their respective learning rates and regularisation parameters (hyper-parameters) to zero or a non-zero value respectively.

Although social relationships are likely to be time-dependent, most datasets do not contain this information. Conditional preferences are related to the feature value preferences, since they model the dependencies between the features and their values, and therefore, are applied to the matrices that account for the users' preferences over feature values. Social influence is applied to the aspects of preferences over features and preferences over feature values. However, applying social influence to the user and item biases showed no observable benefits and user or item biases do not seem to be influenced by social interactions. Therefore, we concluded that user and item biases are not much influenced by the social interactions \citep{zafari2017modelling}. Therefore, in the most abstract view of the model as depicted in the high-level representation in Fig. \ref{fig2a}, the model is comprised of four main modules. Initialising the model parameters (Model Initialiser), learning the intrinsic constituting aspects of preferences (i.e. preferences over features, preferences over feature values, conditional dependencies, and user and item bias values) and the drifting properties of preferences (Intrinsic Trainer), learning the social influence of the friends over the drifting intrinsic preference aspects (Social Trainer), and finally updating the model to reflect the new information extracted from the data about user ratings, time, and social connections (Model Updater). These modules will be discussed in more details later, when we introduce the algorithm in section \ref{Aspect-MF Algorithm}.

\subsubsection{Aspect-MF model formulation}
\label{Aspect-MF}
In this section, we provide the mathematical formulation of the preferences captured in Aspect-MF.
Basically, in Aspect-MF, the user preferences are modelled as a \textit{Bayesian Network} \citep{korb2010bayesian}.
Fig. \ref{fig3} shows the topology or the structure of the Bayesian Network for user preferences that are modelled by Aspect-MF. 

As mentioned earlier, Aspect-MF extends CondTrustFVSVD, by adding the time factor to the aspects of preferences as depicted in Fig \ref{fig1}. In CondTrustFVSVD, the user preferences were captured using the matrices $P$, $Q$, $W$, $Z$, $Y$, $\omega$, $y$, with the hyper-parameters $\sigma_{P}$, $\sigma_{Q}$, $\sigma_{W}$, $\sigma_{Z}$, $\sigma_{\omega}$, $\sigma_{y}$, $\sigma_{Y}$, $\sigma_{bu}$ and $\sigma_{bi}$.

\begin{figure}[!ht]
	\setcounter{figure}{2}
	\hspace*{0.2in}
	\centerline{\includegraphics[scale=0.4]{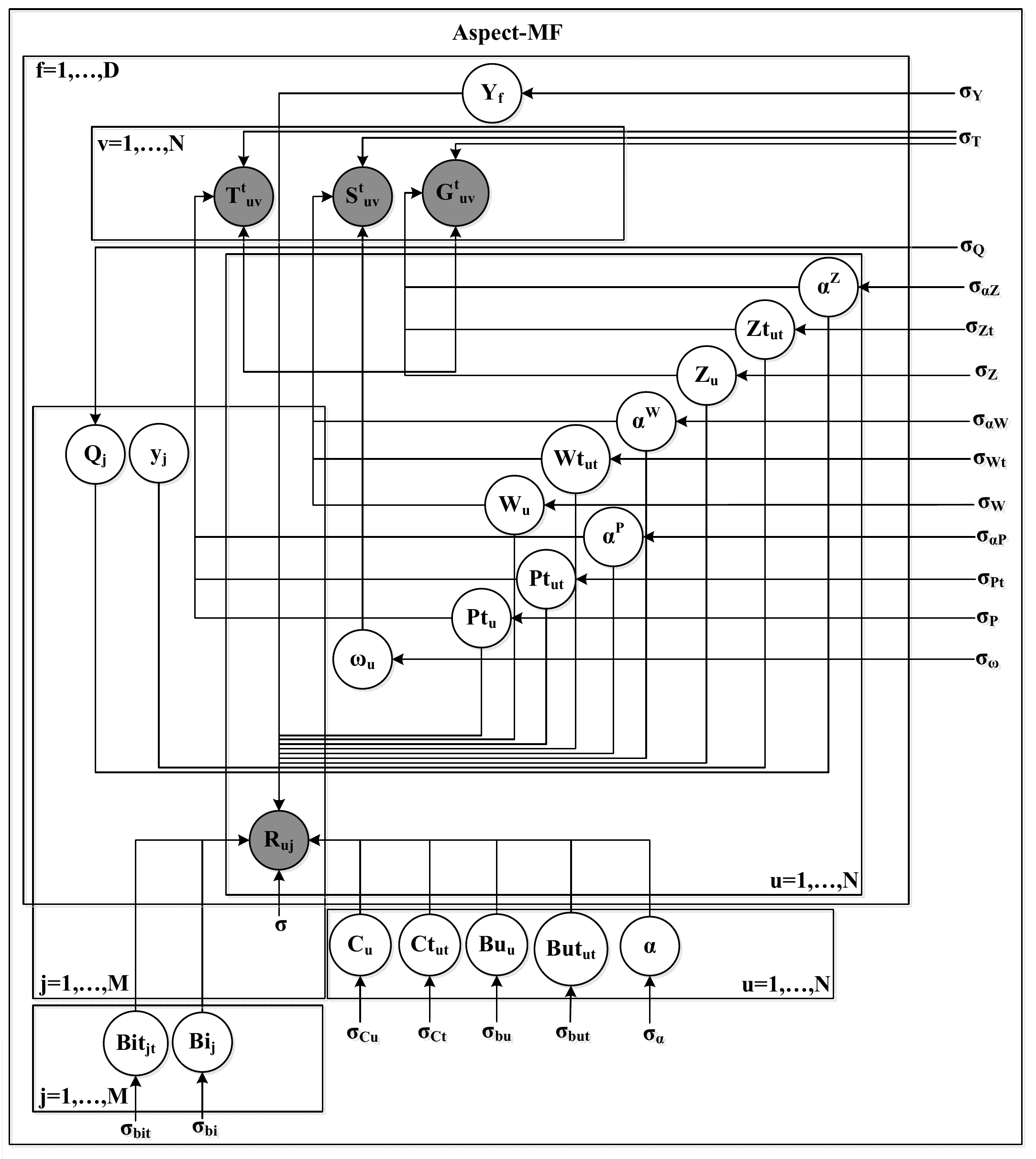}}
	\caption{Bayesian network of Aspect-MF}
	\label{fig3}
\end{figure}

In Aspect-MF, the drifting social influence of friends in the user's social network are captured through Eq. \ref{eq5} to \ref{eq7}.

\begin{equation}
	\begin{split}
		\label{eq5}
		\hat{T}^{t}_{uv} =\frac{1}{|I_u^{t}|} \sum_{\forall {t_{uj}} \in I_u^t}^{D} \sum_{f=1}^{D}P_{uf}({t_{uj}})\omega_{vf}
	\end{split}
\end{equation}

\begin{equation}
	\begin{split}
		\label{eq6}
		\hat{S}^t_{uv} = \frac{1}{|I_u^{t}|} \sum_{\forall {t_{uj}} \in I_u^t}^{D} \sum_{f=1}^{D}(1 - W_{uf}({t_{uj}}))\omega_{vf}
	\end{split}
\end{equation}

\begin{equation}
	\begin{split}
		\label{eq7}
		\hat{G}^t_{uv} = \frac{1}{|I_u^{t}|} \sum_{\forall {t_{uj}} \in I_u^t}^{D} \sum_{f=1}^{D}Z_{uf}({t_{uj}})\omega_{vf}
	\end{split}
\end{equation}

where $\hat{T}^{t}_{uv}$, $\hat{S}^t_{uv}$, $\hat{G}^t_{uv}$ model the time-dependent influence of user $v$ on the preferences of user $u$ for the preferences over features (captured by $P_{uf}(t)$) and preferences over feature values (captured by $W_{uf}(t)$ and $Z_{uf}(t)$), and similar to CondTrustFVSVD, $\omega$ captures \textit{the implicit influence of user $u$ on other users} and is obtained using the matrix factorisation process. As can be seen in Fig. \ref{fig1}, the user preferences over features and feature values in Aspect-MF are subject to social influence, and they also drift over time. In Eqs. \ref{eq5} to \ref{eq7}, $I_u^t$ is the set of timestamps for all the ratings given by user $u$. Therefore, using these equations, the influence of the user $v$ on the preferences of user $u$ is calculated for all the time points, and then it is averaged. 
Intuitively, these equations are telling us that the trust of user $u$ in user $v$ can be estimated by calculating the average of the weighted averages of user $v$'s influence on user $u$'s preferences for different features, in different times. Intuitively, if user $u$ strongly trusts user $v$, his preferences would be more strongly influenced by user $v$. Furthermore, depending on the trust strength of user $u$ in user $v$ and the influence he gets from user $v$ and its direction (positive or negative), the user's preference can be positively or negatively affected. Therefore in Aspect-MF, the user preferences are subject to social influence, and the social influence depends on the strength of their trust in the friends. According to these equations, if there is no relationship between user $u$ and user $v$, user $u$'s preferences will not be directly affected by the social influence of user $v$.

In Aspect-MF, the drifting preference value of the user $u$ over an item $j$ at time $t$ is obtained according to Eq. \ref{eq8}.

\begin{equation}
	\label{eq8}
	\begin{split}
		\hat{R}_{uj}(t_{uj}) = \mu& + bu_u(t_{uj}) + bi_j(t_{uj}) \\&+ \sum_{f=1}^{D}(P_{uf}({t_{uj}}) + |I_u|^{-\frac{1}{2}}\sum_{\forall i \in I_u}y_{if} + |T_u|^{-\frac{1}{2}}\sum_{\forall v \in T_u}\omega_{vf})(W_{uf}({t_{uj}})Q_{jf} + Z_{uf}({t_{uj}})) \\&+  \sum_{f^{'}=1}^{D}(\sum_{f=1}^{D}(W_{uf}({t_{uj}})Q_{jf} + Z_{uf}({t_{uj}}))Y_{ff^{'}})(W_{uf}({t_{uj}})Q_{jf^{'}} + Z_{uf}({t_{uj}}))
	\end{split}
\end{equation}

According to Eq. \ref{eq8}, in Aspect-MF, different aspects of preferences as well as user and item biases are subject to temporal drift.
As can be seen in Eq. \ref{eq5} to \ref{eq8}, the user bias, item bias, preferences over features captured by the matrix $P$, and preferences over feature values captured by the matrices $W$ and $Z$ are subject to temporal drift. In order to model the drifting properties of these aspects, we use Eqs. \ref{eq9} to \ref{eq13}.

\begin{equation}
	\begin{split}
		\label{eq9}
		bu_u(t_{uj}) = bu_u + \alpha_u dev_u(t_{uj}) + but_{ut_{uj}}
	\end{split}
\end{equation}
\begin{equation}
	\begin{split}
		\label{eq10}
		bi_j(t_{uj}) = (bi_j + bi_{jBin(t_{uj})})(C_u + Ct_{u{t_{uj}}})
	\end{split}
\end{equation}
\begin{equation}
	\begin{split}
		\label{eq11}
		P_{uf}({t_{uj}}) = P_{uf} + \alpha_u^P dev_u(t_{uj}) + Pt_{uft_{uj}}
	\end{split}
\end{equation}

\begin{equation}
	\begin{split}
		\label{eq12}
		Z_{uf}({t_{uj}}) = Z_{uf} + \alpha_u^Z dev_u(t_{uj}) + Zt_{uft_{uj}}
	\end{split}
\end{equation}

\begin{equation}
	\label{eq13}
	W_{uf}({t_{uj}}) = W_{uf} + \alpha_u^W dev_u(t_{uj}) + Wt_{uft_{uj}}
\end{equation}

where $P_{uf}$, $W_{uf}$, and $Z_{uf}$ capture the static preferences of the user $u$, while the variables $P_{uft_{uj}}$, $W_{uft_{uj}}$, $Z_{uft_{uj}}$ capture the day-specific variations in the user preferences (e.g. due to the mood of the users in a particular day), and $\alpha_u^P$, $\alpha_u^W$, and $\alpha_u^Z$ model the users' long term preference shifts, and $dev_u(t_{uj})$ is obtained according to Eq. \ref{eq14} \citep{koren2010collaborative}.

\begin{equation}
	\begin{split}
		\label{eq14}
		dev_u(t_{uj}) = sign(t_{u_{uj}} - t_u).{|t_{uj} - t_u|}^\beta
	\end{split}
\end{equation}

where $t_u$ is the mean of the dates for the ratings given by the user $u$, and $\beta$ is a constant value. In Eq. \ref{eq10}, all the dates are placed in a fixed number of bins, and the function $Bin(.)$ returns the bin number for a particular date. For example, if the maximum period of the ratings is 30 years and 30 bins are used, all the rates given in a particular year are placed in a bin, and the function $Bin(.)$ returns the year number for that particular year. The reason why this function is only used for items is that items are not expected to change on a daily basis, and as opposed to users' biases, longer time periods are expected to pass, before we see any changes in the items' popularity.
In simple words, $dev_u(t_{uj})$ shows how much the time of the rating given by user $u$ to the item $j$ deviates from the average time of the ratings given by that user. Therefore, if a rating is given at the same time as the average time of the ratings, then the according to these equations, there will be no long-term preference shift for that aspect. However, for instance, if the average time of the rates given by user $u$ is 11/04/2006, the rating of the same item by that user on 11/04/2016 would be different, and this shift is captured by the coefficients of the function $dev_u(t_{uj})$ in Eq. \ref{eq9} and Eqs. \ref{eq11} to \ref{eq13}. The drifting preferences captured using Eq. \ref{eq9} and Eqs. \ref{eq11} to \ref{eq13} are depicted in Fig. \ref{fig4}. In these figures, the mean of the dates on which the user has given the ratings are assumed to be 50 (the fiftieth day in a year), and the variations of the user preferences over a period of one year are captured for different values of $\alpha$ in Eq. \ref{eq9} and Eqs. \ref{eq11} to \ref{eq13}. The red lines in these figures represent the case in which the day-specific variations in the user preferences are not captured, while the blue lines also include the day-specific variations. Therefore, as can be seen, in these figures there are two types of preference shifts, long term drifts (captured by the values of $\alpha$, $\alpha^P$, $\alpha^W$, and $\alpha^Z$), and short-term or day-specific drifts (captured by the values of $but$, $Pt$, $Wt$, and $Zt$). Therefore, the preference drifts are comprised of small variations from one day to the other, mainly because of temporary factors such as the mood of the user, and the large variations which happen in the long term, as the user changes preferences because of the shift in the his/her tastes. The blue lines show the preference shift patterns that can be learnt by Aspect-MF. Furthermore, the first three terms in Eq. \ref{eq18} model the social influence of the feature preferences and feature value preferences captured by $P$, $\alpha^P$, $Pt$, $W$, $\alpha^W$, $Wt$, $Z$, $\alpha^Z$, $Zt$. Therefore, assuming that two users have established the social relationship from the very beginning (which is not essentially true, but usually social relationships do not contain time-stamps), using the Eqs. \ref{eq5} to \ref{eq7}, the social influence is applied to the preferences of the user over the entire period for which the rating data is record. Therefore, the formulation of the estimated ratings in Aspect-MF (\ref{eq8}) allows it to learn the drifting conditional feature value preferences, and the formulation of the optimisation in Aspect-MF (Eq. \ref{eq18}) enables it to learn the influence of social friends on the drifting preferences of a user.

\begin{figure}[!ht]
	\vskip 0cm
	\setcounter{figure}{3}
	\begin{subfigure}[b]{1\textwidth}
		\includegraphics[height=7cm,width=14cm]{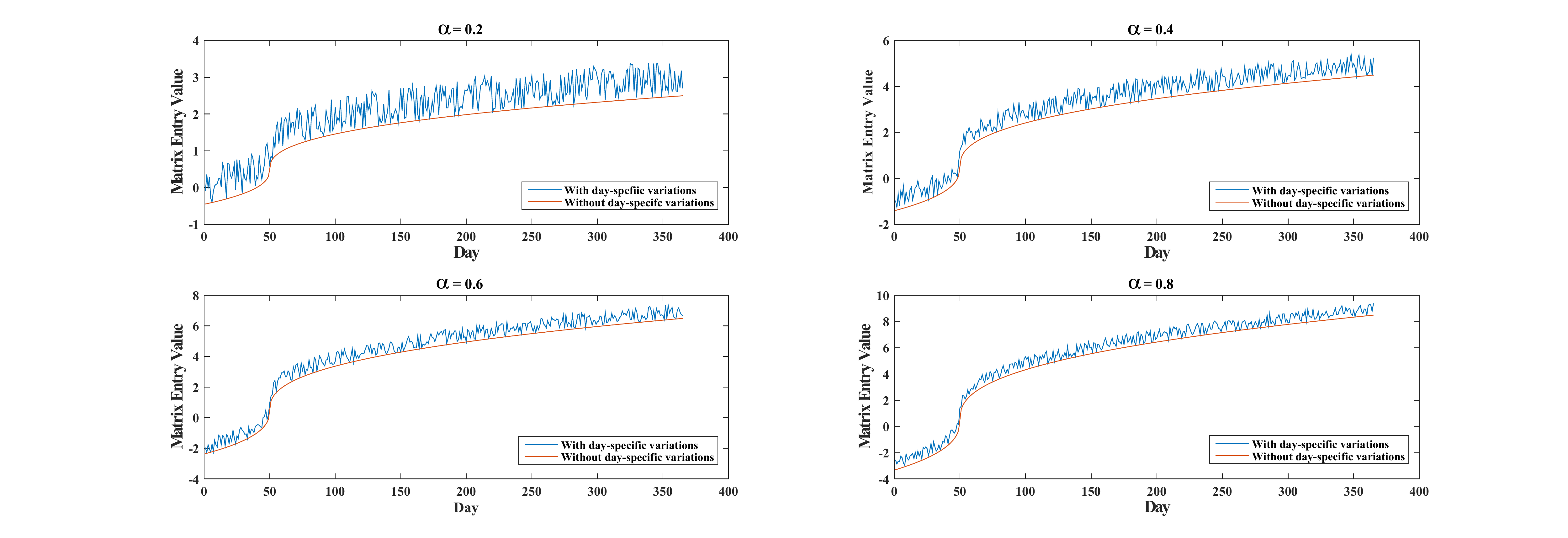}
		\caption{}
		\label{fig4a}
	\end{subfigure}%
	\newline
	\begin{subfigure}[b]{1\textwidth}
		\includegraphics[height=7cm,width=14cm]{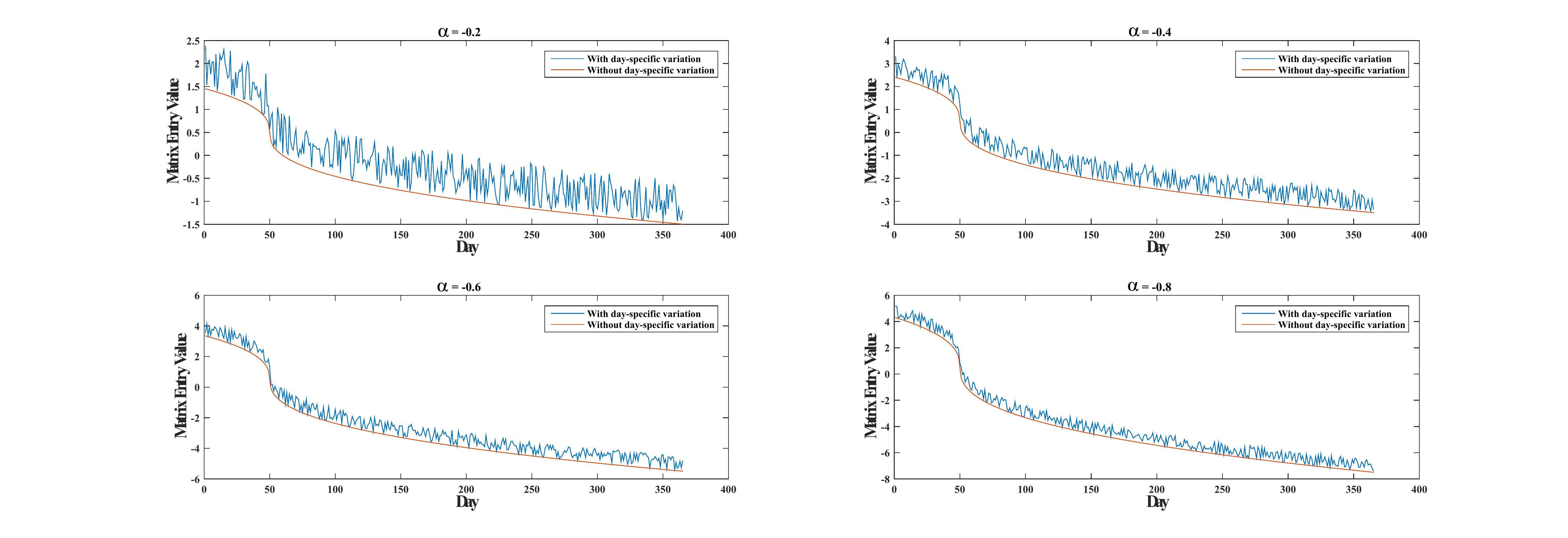}
		\caption{}
		\label{fig4b}
	\end{subfigure}%
	\centering
	\caption{An example of drifting preferences in Eq. \ref{eq9} and Eqs. \ref{eq11} to \ref{eq13} for a) positive $\alpha$ values and b) negative $\alpha$ values}
	\label{fig4}
\end{figure}

Eqs. \ref{eq9} to \ref{eq13} show how Aspect-MF can capture long-term and short-term drifts in each one of the preference aspects (user bias, item bias, feature preferences, and feature value preferences). The advantage of formulating the problem using Eq. \ref{eq8} is that each one these aspects can be arbitrarily switched on/off. This results in a component-based approach, in which the model aspects interact with each other, with the purpose of extracting as much preference patterns from the raw data as possible.

\subsubsection{Aspect-MF model training}
According to the Bayesian network of Aspect-MF in Fig. \ref{fig3}, this model minimises the log-posterior probability of matrices that define the user preferences, given the model hyper-parameters and the training matrix. Formally,

\small
\begin{equation}
	\begin{split}
		\label{eq15}
		&argmin_{P,Pt,\alpha^P,Q,W,Wt,\alpha^W,Z,Zt,\alpha^Z,Y,\omega,y,bu,\alpha,but,C,Ct,bi,bit}\\&\{ln p(P,Q,W,Z,\omega,y,bu,bi,\alpha_u,bu_{t},bi_{Bin(t)},c,c_{t},\alpha^P,\alpha^Z,\alpha^W,P_t,Z_t,W_t\\&|R,T^t,S^t,G^t,\sigma_{N}\}
	\end{split}
\end{equation}
\normalsize
$\sigma_{N}$=\{$\sigma$,$\sigma_{T}$,$\sigma_{P}$,$\sigma_{Pt}$,$\sigma_{\alpha^P}$,$\sigma_{Q}$,$\sigma_{W}$,$\sigma_{Wt}$,$\sigma_{\alpha^W}$,$\sigma_{Z}$,$\sigma_{Zt}$,$\sigma_{\alpha^Z}$,$\sigma_{\omega}$,$\sigma_{y}$,$\sigma_{bu}$,$\sigma_{\alpha}$,$\sigma_{but}$,$\sigma_{C}$,$\sigma_{Ct}$,$\sigma_{bi}$,$\sigma_{bit}$,$\sigma_{Y}$\}\\ denotes the set of all the hyper-parameters. $T^t$, $S^t$, $G^t$ respectively denote the real values for the estimated matrices $\hat{T}^t$, $\hat{S}^t$, and $\hat{G}^t$ in Eqs. \ref{eq5} to \ref{eq7}.
According to the Bayesian network in Figure \ref{fig3} and by decomposing the full joint distribution using chain rule of probability theory \cite{korb2010bayesian} according to the conditional dependencies between the variables defined in this figure, minimising the probability above is equal to minimising the value given in Eq. \ref{eq16} \cite{korb2010bayesian}.

\small
\begin{equation}
	\begin{split}
		\label{eq16}
		&argmin_{P,Pt,\alpha^P,Q,W,Wt,\alpha^W,Z,Zt,\alpha^Z,Y,\omega,y,bu,\alpha,but,C,Ct,bi,bit,Y}\\ \{
		& ln p(R|P(t),Q,W(t),Z(t),bu(t),bi(t),Y,\sigma) 
		+
		ln p(Q|\sigma_{Q}) \\ &
		+
		ln p(P(t)|\sigma_{P})
		+
		ln p(W(t)|\sigma_{W})
		+
		ln p(Z(t)|\sigma_{Z}) \\ &
		+
		ln p(bu(t)|\sigma_{bu})
		+
		ln p(bi(t)|\sigma_{bi}) 
		+
		ln p(y|,\sigma_{y}) 
		+
		ln p(Y|,\sigma_{}) \\ &
		+
		ln p(T^{t}_{uv}|\omega, P(t),\sigma_{T}) 
		+
		ln p(S^{t}_{uv}|\omega, W(t),\sigma_{T})
		+
		ln p(G^{t}_{uv}|\omega, Z(t),\sigma_{T}) \\ &
		+
		ln p(P(t)|\sigma_{T})
		+
		ln p(W(t)|\sigma_{T})
		+
		ln p(Z(t)|\sigma_{T})
		+
		ln p(\omega|,\sigma_{T})
		\}
	\end{split}
\end{equation}
\normalsize

Provided that all the probabilities above follow a normal distribution, it can be shown that minimising the function in Eq. \ref{eq16} is equivalent to minimising the error value using the function in Eq. \ref{eq19}.

\small
\begin{equation}
	\begin{split}
		\label{eq17}
		E_R &=
		\frac{1}{2} \sum_{u=1}^{N}\sum_{j=1}^{M} (\, R_{uj} - \hat{R}_{uj} )\,^2
		+
		\frac{\lambda_{Q}}{2} \sum_{j=1}^{M}\|Q_{j}\|_{Frob}^{2} 
		+
		\frac{\lambda_{y}}{2}\sum_{i=1}^{M}|U_{i}|^{-\frac{1}{2}}\|y_i\|_{Frob}^{2} \\ &
		+
		\sum_{u=1}^{N}\frac{\lambda_P}{2}|I_u|^{-\frac{1}{2}}(\|P_{u}\|_{Frob}^{2} + \|Pt_{ut}\|_{Frob}^{2} + \|\alpha^P\|_{Frob}^{2}) \\ &
		+
		\sum_{u=1}^{N}\frac{\lambda_W}{2}|I_u|^{-\frac{1}{2}}(\|W_{u}\|_{Frob}^{2} + \|Wt_{ut}\|_{Frob}^{2} + \|\alpha^W\|_{Frob}^{2}) \\ &
		+
		\sum_{u=1}^{N}\frac{\lambda_Z}{2}|I_u|^{-\frac{1}{2}}(\|Z_{u}\|_{Frob}^{2} + \|Zt_{ut}\|_{Frob}^{2} + \|\alpha^Z\|_{Frob}^{2}) \\ &
		+
		\sum_{u=1}^{N}\frac{\lambda_Z}{2}|I_u|^{-\frac{1}{2}}(\|Z_{u}\|_{Frob}^{2} + \|Zt_{ut}\|_{Frob}^{2} + \|\alpha^Z\|_{Frob}^{2}) \\ &
		+
		\frac{\lambda_{bu}}{2} \sum_{u=1}^{N}|I_u|^{-\frac{1}{2}}(bu_{u}^{2} + \alpha_{u}^{2} + C_{u}^{2} + \|bu_{u}\|_{Frob}^{2} + \|Ct_{u}\|_{Frob}^{2}) \\ &
		+
		\frac{\lambda_{bi}}{2} \sum_{j=1}^{M}|U_j|^{-\frac{1}{2}}bi_{j}^{2}
		+
		\frac{\lambda_{bi}}{2} \sum_{j=1}^{M}\sum_{\forall t \in I^t_j}^{}|U_j|^{-\frac{1}{2}}bit_{j,Bin(t)}^{2}
		+
		\frac{\lambda_{Y}}{2}\sum_{f=1}^{D}\sum_{f^{'}=1}^{D}Y_{ff^{'}}^2 
	\end{split}
\end{equation}
\normalsize

\small
\begin{equation}
	\begin{split}
		\label{eq18}
		E_T &=
		\frac{\lambda_t\eta_P}{2} \sum_{u=1}^{N}\sum_{\forall v \in T_u} (\, T_{uv} - \hat{T}_{uv} )\,^2 
		+
		\frac{\lambda_t\eta_W}{2} \sum_{u=1}^{N}\sum_{\forall v \in T_u} (\, T_{uv} - \hat{S}_{uv} )\,^2 
		+
		\frac{\lambda_t\eta_Z}{2} \sum_{u=1}^{N}\sum_{\forall v \in T_u} (\, T_{uv} - \hat{G}_{uv} )\,^2 \\ &
		+
		\sum_{u=1}^{N}\frac{\lambda_{T}}{2}|T_u|^{-\frac{1}{2}}(\|P_{u}\|_{Frob}^{2} + \|Pt_{ut}\|_{Frob}^{2} + \|\alpha^P\|_{Frob}^{2}) \\ &
		+
		\sum_{u=1}^{N}\frac{\lambda_{T}}{2}|T_u|^{-\frac{1}{2}}(\|W_{u}\|_{Frob}^{2} + \|Wt_{ut}\|_{Frob}^{2} + \|\alpha^W\|_{Frob}^{2}) \\ &
		+
		\sum_{u=1}^{N}\frac{\lambda_{T}}{2}|T_u|^{-\frac{1}{2}}(\|Z_{u}\|_{Frob}^{2} + \|Zt_{ut}\|_{Frob}^{2} + \|\alpha^Z\|_{Frob}^{2}) \\ &
		+
		\frac{\lambda_{\omega}}{2} \sum_{v=1}^{N}|T^{+}_{v}|^{-\frac{1}{2}}\|\omega_v\|_{Frob}^{2} \\ &
	\end{split}
\end{equation}
\normalsize

\begin{equation}
	\begin{split}
		\label{eq19}
		argmin_{P,Pt,\alpha^P,Q,W,Wt,\alpha^W,Z,Zt,\alpha^Z,Y,\omega,y,bu,\alpha,but,C,Ct,bi,bit}
		[ \,
		E = E_R + E_T
		] \,
	\end{split}
\end{equation}

where $I_j^t$ is the set of timestamps, for all the ratings given to item $j$, and $\eta_P$, $\eta_W$, and $\eta_Z$ are constants added to control the weights of the components related to the social aspect in this equation. The details of the model training can be found in Appendix \ref{Aspect-MF Training Equations}.

\subsubsection{Aspect-MF algorithm}
\label{Aspect-MF Algorithm}
Algorithm \ref{ModelTrainer} describes the details of the gradient descent method Aspect-MF uses to train the model parameters ($P$, $Pt$, $\alpha^P$, $Q$, $W$, $Wt$, $\alpha^W$, $Z$, $Zt$, $\alpha^Z$, $Y$, $\omega$, $y$, $bu$, $\alpha$, $but$, $C$, $Ct$, $bi$, $bit$) as expressed in Eq. \ref{eq19}.

The algorithm receives the set of model hyper-parameters $\lambda$ and the set of learning rates $\gamma$ as input, and trains the model parameters according to the Bayesian approach described in section \ref{Aspect-MF}. 
As we showed in the high-level representation of the algorithm in Figure \ref{fig2a}, the model is comprised of four basic components. A model initialiser, which initialises the model parameters after the input data is loaded into memory, an intrinsic trainer, which trains the model parameters using the user-item ratings, a social trainer which trains the model parameters using the social relationship data, and finally, a model updater, which updates the model based on the trained parameters for a particular iteration.

As can be seen in line \ref{Init} in Algorithm \ref{ModelTrainer}, the training starts with initialising the model parameters. The matrices $P$, $Q$, $y$, and $\omega$ and user and item bias vectors ($bu$ and $bi$) are randomly initialised using a Gaussian distribution with a mean of zero and the standard deviation of one. The new matrices $Pt$, $W$, $Wt$, $Z$, $Zt$, $Ct$, $but$, $bit$, and $Y$ and the vectors $\alpha$, $\alpha^P$, $\alpha^W$, $\alpha^Z$, $C$ are initialised with constant values.
By using constant values to initialise the matrices and vectors, the algorithm starts the search process at the same starting point as CTFVSVD, and explores the modified search space to find more promising solutions, by considering the possible conditional dependencies between the features and the differences between users in preferring item feature values, as well as dynamic properties of the preferences, and the influence of social friends in the preferences of a user.

The main algorithm consists of a main loop, which implements the learning iterations of the model. Each iteration is comprised of one model intrinsic training operation (Algorithm \ref{IntrinsicTrainer}), one model social training operation (Algorithm \ref{SocialTrainer}), and one model updating operation (Algorithm \ref{ModelUpdater}).
In the model intrinsic trainer, the model parameters are updated using the gradient values in Eqs. \ref{eqa1} to \ref{eqa41}, using a rating value that is read from the user-item ratings matrix. 
First in line \ref{estimatedrating}, the estimated rating is calculated according to Eq. \ref{eq8}. 
Then the basic parameters of the model, $P$, $Q$, $W$, $Z$, $Y$, $bu$, and $bi$, and the temporal parameters $but$, $bit$, $\alpha$, $C$, $Ct$, $\alpha^P$, $\alpha^W$, $\alpha^Z$, $Pt$, $Wt$, and $Zt$ are updated using the rating-related gradient values ($\frac{\partial E_R}{\partial(.)}$) in the Eqs. \ref{eqa1} to \ref{eqa41}. Since this trainer only learns the intrinsic user preferences, only the error value in Eq. \ref{eq17} will be used to update the model parameters. After learning the intrinsic preferences, the function in Algorithm \ref{SocialTrainer} is invoked to train the social aspects of the preferences. Similar to IntrinsicTrainer, SocialTrainer is also comprised of a main loop, which iterates over the social relationship data in the social matrix. In each iteration, one entry from the social matrix is read, and the  socially-influenced parameters of the model are updated though the gradient values that are obtained using the error in Eq. \ref{eq18}. Finally, the ModelUpdater in Algorithm \ref{ModelUpdater} is invoked, and the calculated model updates are applied to the model parameters. This process is repeated for a fixed number of iterations, or until a specific condition is met. At the end of this process, the model parameters ($P$, $Pt$, $\alpha^P$, $Q$, $W$, $Wt$, $\alpha^W$, $Z$, $Zt$, $\alpha^Z$, $Y$, $\omega$, $y$, $bu$, $\alpha$, $but$, $C$, $Ct$, $bi$, $bit$) are trained using the input data, and can be used to estimate the rating value given by a user $u$ to an item $j$ according to Eq. \ref{eq8}.

\begin{tcolorbox}[blanker,float=bpt, grow to left by=1cm, grow to right by=1cm]
	
	\begin{algorithm}[H]
		\caption{Model Training}\label{ModelTrainer}
		\begin{algorithmic}[1]
			\State{\textbf{void} ModelTrainer($\lambda$, $\gamma$, $maxIter$)}
			\footnote{$\lambda$ is the set of the model hyper-parameters as specified in Eqs. \ref{eq17} and \ref{eq18} and Figure \ref{fig1}.
				$N$, $M$, and $D$ respectively denote number of users, number of items, and number of features.
				$\gamma$ denotes the set of learning rates, $maxIter$ denotes the maximum number of learning iterations.}
			\State {$\lambda$ =\{$\lambda_{T}$,$\lambda_{P}$,$\lambda_{Pt}$,$\lambda_{\alpha^P}$,$\lambda_{Q}$,$\lambda_{W}$,$\lambda_{Wt}$,$\lambda_{\alpha^W}$,$\lambda_{Z}$,$\lambda_{Zt}$,$\lambda_{\alpha^Z}$,$\lambda_{\omega}$,$\lambda_{y}$,$\lambda_{bu}$,$\lambda_{\alpha}$,$\lambda_{but}$,$\lambda_{C}$,$\lambda_{Ct}$,$\lambda_{bi}$,$\lambda_{bit}$,$\lambda_{Y}$\}}
			\State {$\gamma$ =\{$\gamma_{T}$,$\gamma_{P}$,$\gamma_{Pt}$,$\gamma_{\alpha^P}$,$\gamma_{Q}$,$\gamma_{W}$,$\gamma_{Wt}$,$\gamma_{\alpha^W}$,$\gamma_{Z}$,$\gamma_{Zt}$,$\gamma_{\alpha^Z}$,$\gamma_{\omega}$,$\gamma_{y}$,$\gamma_{bu}$,$\gamma_{\alpha}$,$\gamma_{but}$,$\gamma_{C}$,$\gamma_{Ct}$,$\gamma_{bi}$,$\gamma_{bit}$,$\gamma_{Y}$\}}
			\State $\{$
			\State {\textit{//Creating matrices $P$, $\omega$, $W$, and $Z$ and temporary matrices $P^S$, $\omega^S$, $W^S$, and $Z^S$:}}
			\State 
			$Matrix $ $P,P^{S};$
			$Matrix $ $\omega^{S};$
			$Matrix $ $W^{S};$
			$Matrix $ $Z^{S};$
			\State {\textit{//Creating vectors $\alpha^{P}$, $\alpha^{W}$, and $\alpha^{W}$, and temporary vectors $\beta^{P}$, $\beta^{W}$, and $\beta^{W}$:}}
			\State 
			$Vector$ $\alpha^{P},\beta^{P};$
			$Vector $ $\alpha^{W},\beta^{W};$
			$Vector $ $\alpha^{Z},\beta^{Z};$
			\State {\textit{//Creating tables $Pt$, $Wt$, and $Zt$, and temporary tables $Pt^S$, $Wt^S$, and $Zt^S$:}}
			\State 
			$Table $ $Pt,Pt^{S};$
			$Table $ $Wt,Wt^{S};$
			$Table $ $Zt,Zt^{S};$
			\State {ModelInitialiser();}\label{Init}
			
			\State $\textit{l} \gets \textit{1};$
			
			\For{\textit{l} $\preceq$ $maxIter$}\label{learningLoop}
			
			\State $IntrinsicTrainer();$
			
			\State $SocialTrainer();$
			
			\State $ModelUpdater();$
			
			\State {$error \gets error \times 0.5;$}
			\State $\textit{l} \gets \textit{l} + 1;$
			\EndFor
			
			\State $\}$
		\end{algorithmic}
	\end{algorithm}
	\begin{algorithm}[H]
		\caption{Model Initialising}\label{ModelInitialiser}
		\begin{algorithmic}[1]
			\State {\textbf{void} ModelInitialiser($\lambda$, $\gamma$)}
			\State $\{$
			\State $initMean \gets 0; initStd \gets 1;$
			\State $P.init(initMean, initStd); \alpha^P.initConst(0); Pt.initConst(0);$
			\State $P^S.init(initMean, initStd); \beta^P.initConst(0); Pt^S.initConst(0);$
			\State $W.initConst(0); \alpha^W.initConst(0); Wt.initConst(0);$
			\State $W^S.init(initMean, initStd); \beta^W.initConst(0); Wt^S.initConst(0);$
			\State $Z.initConst(0); \alpha^Z.initConst(0); Zt.initConst(0);$
			\State $Z^S.init(initMean, initStd); \beta^Z.initConst(0); Zt^S.initConst(0);$
			\State $\omega.init(initMean, initStd); \omega^S.init(initMean, initStd);$
			\State $bu.init(initMean, initStd); \alpha.init(0); but.init(0); C.init(0); Ct.init(0);$
			\State $bi.init(initMean, initStd); \beta.initConst(0); bit.initConst(0);$
			\State $Q.init(initMean, initStd); y.init(initMean, initStd);$
			\footnote{$initMean$ and $initStd$ are the mean and standard deviation values that are used to initialise the model parameters.
				init(initMean, initStd) is a function that initialises a bias vector (e.g. $bu$ and $bi$) and a matrix (e.g. $P$, and $Q$) using Gaussian distribution with mean value of $initMean$ and standard deviation of $initStd$.
				initConst(initMean, initStd) initialises a matrix (e.g. $W$ and $Z$) with a constant value.}
			\State $\}$
		\end{algorithmic}
	\end{algorithm}
\end{tcolorbox}

\begin{tcolorbox}[blanker,float=bpt, grow to left by=1cm, grow to right by=1cm]
	\begin{algorithm}[H]
		\caption{Intrinsic Training}\label{IntrinsicTrainer}
		\begin{algorithmic}[1]
			\State {\textbf{void} IntrinsicTrainer($\lambda$, $\gamma$)}
			\State $\{$
			\State $\textit{u} \gets \textit{1};$
			\For {\textit{u} $\preceq$ \textit{N}} \label{forloop1}
			\State $\textit{j} \gets \textit{1};$
			\For {\textit{j} $\preceq$ \textit{M}}
			
			\If {$R_{uj} \ne 0$}
			
			\State Calculate $\hat{R}_{uj}$ according to Eq. \ref{eq8}.\label{estimatedrating}
			
			\State Get the time $t$ that the rating $R_{uj}$ has been given.
			
			\State {Update $bu_{u}$, $but_{ut}$, and $\alpha_{u}$ according to Eqs. \ref{eqa1}-\ref{eqa3} using $\gamma_\alpha$, $\gamma_{bu}$, $\gamma_{but}$;}\label{updateRulebu}
			
			\State {Update $bi_{j}$ and $bit_{jt}$ according to Eqs. \ref{eqa4}-\ref{eqa5} using $\gamma_{bi}$ and $\gamma_{bit}$;}\label{updateRulebi}
			
			\State {Update $C_{u}$ and $Ct_{ut}$ according to Eqs. \ref{eqa6}-\ref{eqa7} using  $\gamma_{C}$ and $\gamma_{Ct}$;}\label{updateRuleC}
			
			\State $\textit{f} \gets \textit{1};$
			\For{\textit{f} $\preceq$ \textit{D}}
			
			\State {Update $P^S_{uf}$, $Pt^S_{uft}$, and $\beta^{P}_{u}$ according to Eqs. \ref{eqa9}, \ref{eqa12}, and \ref{eqa15} using $\gamma_P$, ${\gamma_{Pt}}$, and $\gamma_{\alpha^P}$;}\label{updateRuleP1}		
			
			\State {Update $Q_{jf}$ according to Eq. \ref{eqa40} using $\gamma_Q$;}\label{updateRuleQ}
			
			\State {Update $W^S_{uf}$, $Wt^S_{uft}$, and $\beta^{W}_{u}$ according to Eqs. \ref{eqa18}, \ref{eqa21}, and \ref{eqa24} using $\gamma_W$, ${\gamma_{Wt}}$, and $\gamma_{\alpha^W}$;}\label{updateRuleW1}
			
			\State {Update $Z^S_{uf}$, $Zt^S_{uft}$, and $\beta^{Z}_{u}$ according to Eqs. \ref{eqa27}, \ref{eqa30}, and \ref{eqa33} using $\gamma_Z$, ${\gamma_{Zt}}$, and $\gamma_{\alpha^Z}$;}\label{updateRuleZ1}
			
			\State {$\forall v \in T_{u}$: Update $\omega^S_{vf}$ according to Eq. \ref{eqa35} using $\gamma_\omega$;}\label{updateRuleOmega1}
			\State {$\forall i \in I_{u}$: Update $y_{if}$ according to Eq. \ref{eqa33} using $\gamma_y$;}\label{updateRuley}
			
			\State $f^{'} \gets \textit{f}+1;$
			\For{$f^{'}$ $\preceq$ $\textit{D}$}
			
			\State {Update $Y_{ff^{'}}$ and $Y_{f^{'}f}$ according to Eq. \ref{eqa39} using $\gamma_Y$;} \label{updateRuleY}			
			
			\State $f^{'} \gets f^{'} + 1;$
			\EndFor
			
			\State $\textit{f} \gets \textit{f} + 1;$
			\EndFor
			
			\EndIf
			
			\State $\textit{j} \gets \textit{j} + 1;$
			\EndFor
			\State $\textit{u} \gets \textit{u} + 1;$
			\EndFor
			\State $\}$
		\end{algorithmic}
	\end{algorithm}
\end{tcolorbox}

\begin{tcolorbox}[blanker,float=bpt, grow to left by=1cm, grow to right by=1cm]
	\begin{algorithm}[H]
		\caption{Social Training}\label{SocialTrainer}
		\begin{algorithmic}[1]
			\State {\textbf{void} SocialTrainer($\lambda$, $\gamma$)}
			\State $\{$
			\State $\textit{u} \gets \textit{1};$
			\For {\textit{u} $\preceq$ \textit{N}}\label{forloop2}
			\State $\textit{v} \gets \textit{1};$
			\For {\textit{v} $\preceq$ \textit{N}}
			
			\If {\textit{v} $\in$ \textit{$T_u$}}		
			
			\For{\textit{f} $\preceq$ \textit{D}}
			
			\State {Update $P^S_{uf}$, $W^S_{uf}$, and $Z^S_{uf}$ according to Eqs. \ref{eqa10}, \ref{eqa19}, \ref{eqa28} using $\gamma_P$, $\gamma_W$, and $\gamma_Z$;}\label{updateRulePWZ2}
			
			\State {$\forall t \in I_u^t:$ Update $Pt^S_{uft}$, $Wt^S_{uft}$, and $Zt^S_{uft}$ according to Eqs. \ref{eqa13}, \ref{eqa16}, \ref{eqa19} using $\gamma_{Pt}$, $\gamma_{Wt}$, and $\gamma_{Zt}$;}\label{updateRulePWZ3}
			
			\State {Update $\beta^P_{uf}$, $\beta^W_{uf}$, and $\beta^Z_{uf}$ according to Eqs. \ref{eqa16}, \ref{eqa19}, \ref{eqa22} using $\gamma_{\alpha^P}$, $\gamma_{\alpha^W}$, and $\gamma_{\alpha^Z}$;}\label{updateRuleAlphaPWZ2}
			\State {$\forall t \in I_u^t:$ Update $\omega^t_{vf}$ according to Eq. \ref{eqa26} using $\gamma_\omega$;}

			\State $\textit{f} \gets \textit{f} + 1;$
			\EndFor
			
			\EndIf
			
			\State $\textit{v} \gets \textit{v} + 1;$
			\EndFor
			\State $\textit{u} \gets \textit{u} + 1;$
			\EndFor
			\State $\}$
		\end{algorithmic}
	\end{algorithm}
	\begin{algorithm}[H]
		\caption{Model Updating}\label{ModelUpdater}
		\begin{algorithmic}[1]
			\State {\textbf{void} ModelUpdater($\lambda$, $\gamma$)}
			\State $\{$
			\State 
			{$\forall u,f: P_{uf} \gets - \gamma_{U} \times P^{S}_{uf};$}
			\State
			{$\forall u: \alpha^P_u \gets - \gamma_{\alpha^P} \times \beta^P_u;$}
			\State
			{$\forall u,f: W_{uf} \gets - \gamma_{W} \times W^{S}_{uf};$}
			\State
			{$\forall u: \alpha^W_u \gets - \gamma_{\alpha^W} \times \beta^W_u;$}
			\State
			{$\forall u,f: Z_{uf} \gets - \gamma_{Z} \times Z^{S}_{uf};$}
			\State
			{$\forall u: \alpha^Z_u \gets - \gamma_{\alpha^Z} \times \beta^Z_u;$}
			\State
			{$\forall u,f: \omega_{uf} \gets - \gamma_{\omega} \times \omega^{S}_{uf};$}
			\State $\}$
		\end{algorithmic}
	\end{algorithm}
\end{tcolorbox}

\subsubsection{Computational complexity analysis}
\label{Computational Complexity Analysis}
The model training in Algorithm \ref{ModelTrainer} is comprised of one main loop that iterates for a fixed number of iterations (maxIter). Therefore, the computation time of the model trainer is expressed in Eq. \ref{eq41}.

\small
\begin{equation}
	\label{eq41}
	\begin{split}
		C(ModelTrainer) & = C(IntrinsicTrainer) + C(SocialTrainer) + C(ModelUpdater)
	\end{split}
\end{equation}
\normalsize

First, we examine the computational complexity of Intrinsic Training in Algorithm \ref{IntrinsicTrainer}. On the highest level, this algorithm is comprised of two loops that iterate over the non-zero ratings in the rating matrix $R$. In the following, $|R|$ and $|T|$ denote the number of non-zero entries in the rating matrix $R$ and adjacency matrix $T$ respectively. In Intrinsic Trainer:

\begin{itemize}
	\item The number of repetitions to calculate the estimated ratings ($\hat{R}$) in line \ref{estimatedrating} is $(D^2\times|R|) + (D\times\sum_{u=1}^{N}|I_u|^2) + (D\times\sum_{u=1}^{N}|I_u|\times|T_u|)$.
	\item The number of repetitions to update parameters related to user and item biases in lines \ref{updateRulebu}, \ref{updateRulebi}, and \ref{updateRuleC} is $7\times|R|$.
	\item The number of repetitions needed to update the parameters $P$, $Q$, $W$, and $Z$ in lines \ref{updateRuleP1}, \ref{updateRuleQ}, \ref{updateRuleW1}, and \ref{updateRuleZ1} is $10\times D\times|R|$.
	\item The number of repetitions needed to update the parameters $\omega$ in line \ref{updateRuleOmega1} is $D\times\sum_{u=1}^{N}(|I_u|\times|T_u|)$.
	\item The number of repetitions needed to update the parameters $y$ in line \ref{updateRuley} is $D\times\sum_{u=1}^{N}|I_u|^2$.
	\item The number of repetitions needed to update the dependency matrix $Y$ in line \ref{updateRuleY} is $D^2\times|R|$.
\end{itemize}

Therefore, the overall number of repetitions for the Intrinsic Trainer is obtained according to Eq. \ref{eq42}.

\small
\begin{equation}
	\label{eq42}
	\begin{split}
		N(IntrinsicTrainer)&= D^2\times|R| + D\times \sum_{u=1}^{N}|I_u|\times|T_u|
		+ 7\times|R|
		+ 10\times D\times|R|
		\\ &
		+ D\times\sum_{u=1}^{N}(|I_u|\times|T_u|)
		+ D\times\sum_{u=1}^{N}|I_u|^2
		\\ &
		+ D^2\times|R|
	\end{split}
\end{equation}
\normalsize

Assuming that on average, each user rates $c$ items, and trusts $k$ users, the computation time can be obtained as Eq. \ref{eq43}.

\small
\begin{equation}
	\label{eq43}
	\begin{split}
		C(IntrinsicTrainer) = O(D^2\times|R|) + O(D\times c\times |R|) + O(D\times k\times |T|)
	\end{split}
\end{equation}
\normalsize

Assuming that $c,k \lll N$, we can ignore the values of $c$ and $k$. Therefore, the computational time of the Intrinsic Trainer would be obtained according to Eq. \ref{eq44}.

\small
\begin{equation}
	\label{eq44}
	\begin{split}
		C(IntrinsicTrainer) = O(D^2\times|R|) + O(D\times|R|) + O(D\times |T|) = O(D^2\times|R|) + O(D\times |T|)
	\end{split}
\end{equation}
\normalsize

Consequently, the overall computation time is linear with respect to the number of observed ratings as well as observed trust statements. 
Social Trainer consists of two loops that iterate over the non-zero trust relations in the adjacency matrix $T$. The number of repetitions needed to update the parameters $P$, $W$, $Z$, and $\beta^P$, $\beta^W$, and $\beta^Z$ is $6\times D\times|T|$. The number of repetitions to update the values of $Pt$, $Wt$, $Zt$, and $\omega$ is equal to $4\times(\sum_{u=1}^{N}|I_u|\times|T_u|\times D)$. Therefore, the computation time of Social Trainer is equal to:

\small
\begin{equation}
	\label{eq45}
	\begin{split}
		C(IntrinsicTrainer) = O(D\times |R|) + O(D\times |T|)
	\end{split}
\end{equation}
\normalsize

In the Model Updater, the values of matrices $P$, $W$, $Z$, and vectors $\omega$, $\alpha^P$, $\alpha^W$, and $\alpha^Z$ need to be updated. The computation time needed to update these parameters is $O(N\times D)$. Assuming that each user has rated at least one item, it is safe to say that $|R|$ is greater than the number of users $N$.
Therefore, the computation time of Model Updater does not exceed the maximum computation time of Intrinsic Trainer and Social Trainer. Finally, the computation time of the Model trainer is obtained as Eq. \ref{eq46}.

\small
\begin{equation}
	\label{eq46}
	\begin{split}
		C(ModelTrainer) = O(D^2\times |R|) + O(D\times |T|)
	\end{split}
\end{equation}
\normalsize

The number of latent factors $D$ is fixed, hence the computation time is only a function of $|R|$ and $|T|$. Since both ratings matrix and social network matrix are sparse, the algorithm is scalable to the problems with millions of users and items.

\section{Experiments}
\label{Experiments}

\subsection{Datasets}
We tested Aspect-MF on three popular datasets, Ciao, Epinions, and Flixster.
Ciao is a dataset crawled from the ciao.co.uk website. This dataset includes 35,835 ratings given by 2,248 users over 16,861 movies. Ciao also includes the trust relationships between users. The number of trust relationships in Ciao is 57,544. Therefore the dataset density of ratings and trust relationships are 0.09\% and 1.14\% respectively. The ratings are integer values between 1 and 6.
The Epinions dataset consists of 664,824 ratings from 40,163 users on 139,738 items of different types (software, music, television show, hardware, office appliances, ...). Ratings are integer values between 1 and 5, and data density is 0.011\%. Epinions also enables the users to issue explicit trust statements about other users. This dataset includes 487183 trust ratings. The density of the trust network is 0.03\%.
Flixster is a social movie site which allows users to rate movies and share the ratings with each other, and become friends with others with similar movie taste. The Flixster dataset which is collected from the Flixster website includes 8,196,077 ratings issued by 147,612 users on 48,794 movies. The social network also includes 7,058,819 friendship links. The density of the ratings matrix and social network matrix are 0.11\% and 0.001\% respectively.

In all the experiments in sections \ref{Discussion}, \ref{Statistical Analysis}, and \ref{Dynamic aspects}, 80\% of the datasets are used for training and the remaining 20\% are used for evaluation. In order to achieve statistical significance, each model training is repeated for 30 times and the average values are used. In section \ref{Effect of the size of the training dataset}, we analyse the behaviour of the models in other cases, where 60\% and 40\% of the ratings are used for training.

\subsection{Comparisons}
\label{Comparisons}
In order to show the effectiveness of Aspect-MF, we compared the results against the recommendation quality of some of the most popular state of the art models that have reported the highest accuracies in the literature. The following models are compared across the experiments in this section:
\begin{itemize}
	\item \textit{TrustSVD} \citep{guo2015trustsvd}, which builds on SVD++ \citep{koren2011advances}. The missing ratings are calculated based on explicit and implicit feedback from user ratings and user's trust relations.
	\item \textit{CondTrustFVSVD} \citep{zafari2017modelling}, this method extends TrustSVD by adding the conditional preferences over feature values to TrustSVD. Experimental results show that this method is significantly superior to TrustSVD in terms of accuracy. This model is denoted CTFVSVD in the experiments section.
	\item \textit{Aspect-MF}, which is the model proposed in this paper. The component-based approach that we took in designing this model enabled us to arbitrarily switch on/off the dynamicity over different preference aspects. Therefore, in the experiments we try all the combinations of dynamic preference aspects. This results in 7 combinations denoted by $b$, $bf$, $bffv$, $bfv$, $f$, $ffv$, and $fv$ \footnote{\textit{fv} denotes feature value preferences, \textit{f} denotes feature preferences, and \textit{b} denotes bias. Therefore, \textit{bffv} denotes a model with all the three aspects.}.
\end{itemize}

Guo, Zhang and Yorke-Smith \citep{guo2016novel} carried out comprehensive experiments, and showed that their model, TrustSVD outperformed all the state of the art models. Recently, Zafari and Moser \citep{zafari2017modelling} showed that their model CondTrustFVSVD significantly outperforms TrustSVD. Therefore, in this section, we limited our comparisons to these two models from the state of the art, since they outperform a comprehensive set of state of the art recommendation models \citep{guo2016novel,zafari2017modelling}. 

The optimal experimental settings for each method are determined either by our experiments or suggested by previous works \citep{guo2015trustsvd,guo2016novel,zafari2017proposing}.
Due to the over-fitting problem, the accuracy of iterative models improves for a  number of iterations, after which it starts to degrade. Therefore, we recorded the best accuracy values achieved by each model during the iterations, and compared the models based on the recorded values. We believe that this approach results in a fairer comparison of the models than setting the number of iterations to a fixed value, because the models over-fit at different iterations, and using a fixed number of iterations actually prevents us from fairly comparing the models based on their real capacity in uncovering hidden patterns from data. Therefore, the reported results for iterative models here are the best results that they could achieve using the aforementioned parameters. MAE and RMSE measures are used to evaluate and compare the accuracy of the models.
MAE and RMSE are two standard and popular measures that are used to measure and compare the performance of preference modelling methods in recommender systems. In the following sections, we consider the performances separately for All Users and Cold-start Users. Cold-start Users are the users who have rated less than 5 items, and All Users include all the users regardless of the number of items they have rated.

\subsection{Discussion}
\label{Discussion}
All latent factor approaches have been evaluated with 5 factors, because no clear ideal value could be established. In section \ref{Model Performances}, first we analyse the performance of the models from different perspectives. Since the results are subject to randomness, we also performed a t test to guarantee that the out-performances achieved do not happen by chance. The results are discussed in section \ref{Statistical Analysis}. As we mentioned in section \ref{Introduction}, one of the research questions we are interested in, in this paper is related to the interplay between the dynamicity of preference aspects and the preference domain. In section \ref{Dynamic aspects}, we consider the performance of combinations of Aspect-MF, in order to pinpoint the aspects that are more subject to temporal drift in each dataset. In section \ref{Effect of the size of the training dataset}, we also consider the effect of the amount of training data that is fed to the model as input, and analyse the robustness of the models to the shortage of training data.

\subsubsection{Model performances}
\label{Model Performances}
We can consider the performance of the models from different perspectives. 
A preference model's performance can be considered with respect to the dataset on which it is trained, the accuracy measure that is used to evaluate the model's performance, and the performance of the model on cold-start users vs the performance on all users.

\paragraph{\textbf{Datasets}}
\label{Datasets}
The error values in Fig. \ref{fig5} show that the Aspect-MF results in substantial improvements over TrustSVD in all three datasets for both measures and for all users and cold-start users. As we can see in this figure, the box plots of Aspect-MF's combinations do not have much overlap with the box plot of TrustSVD, which means that the differences are definitely statistically significant. In this figure, we can also see that the box plot widths for Aspect-MF's combinations are usually much smaller than that for TrustSVD. This suggests that Aspect-MF's combinations are more stable than TrustSVD, meaning that they find roughly the same solutions across different model executions. This is a favourable property of the model, since it makes the model performance less subject to randomness. Clearly, a model that performs well sometimes and worse at other times is less reliable. The model's superior performance is likely due to its taking multiple preference aspects into account, therefore, it has more clues as to where the optimal solutions might reside in the solution space.

In particular, we can see that the model is more stable in the case of the Ciao and Epinions datasets than the Flixster dataset. On the Epinions dataset, each typical user and cold-start user rates 41.61 items and 4.08 items on average. These numbers respectively are 15.94 and 2.94 for the Ciao dataset, and 11.12 and 1.94 for the Flixster dataset. This could explain why the variations are larger on Flixster dataset than Epinions and Ciao datasets. Since more ratings per user are available in the Ciao and Epinions dataset, different executions lead the model to more similar solutions than the solutions that are found on the Flixster dataset across different model executions. 
We can also see from Table \ref{table1}, that on the Ciao and Flixster datasets, the improvements are more significant for RMSE, while more significant improvements are achieved for RMSE.
We can also clearly observe that the model variations are smaller for all users in the Epinions dataset, and for cold-start users in the Flixster dataset.

\begin{figure}[!ht]
	\vskip 0cm
	\centering
	\setcounter{figure}{4}
	\begin{subfigure}[b]{0.5\textwidth}
		\includegraphics[height=5cm,width=6.5cm]{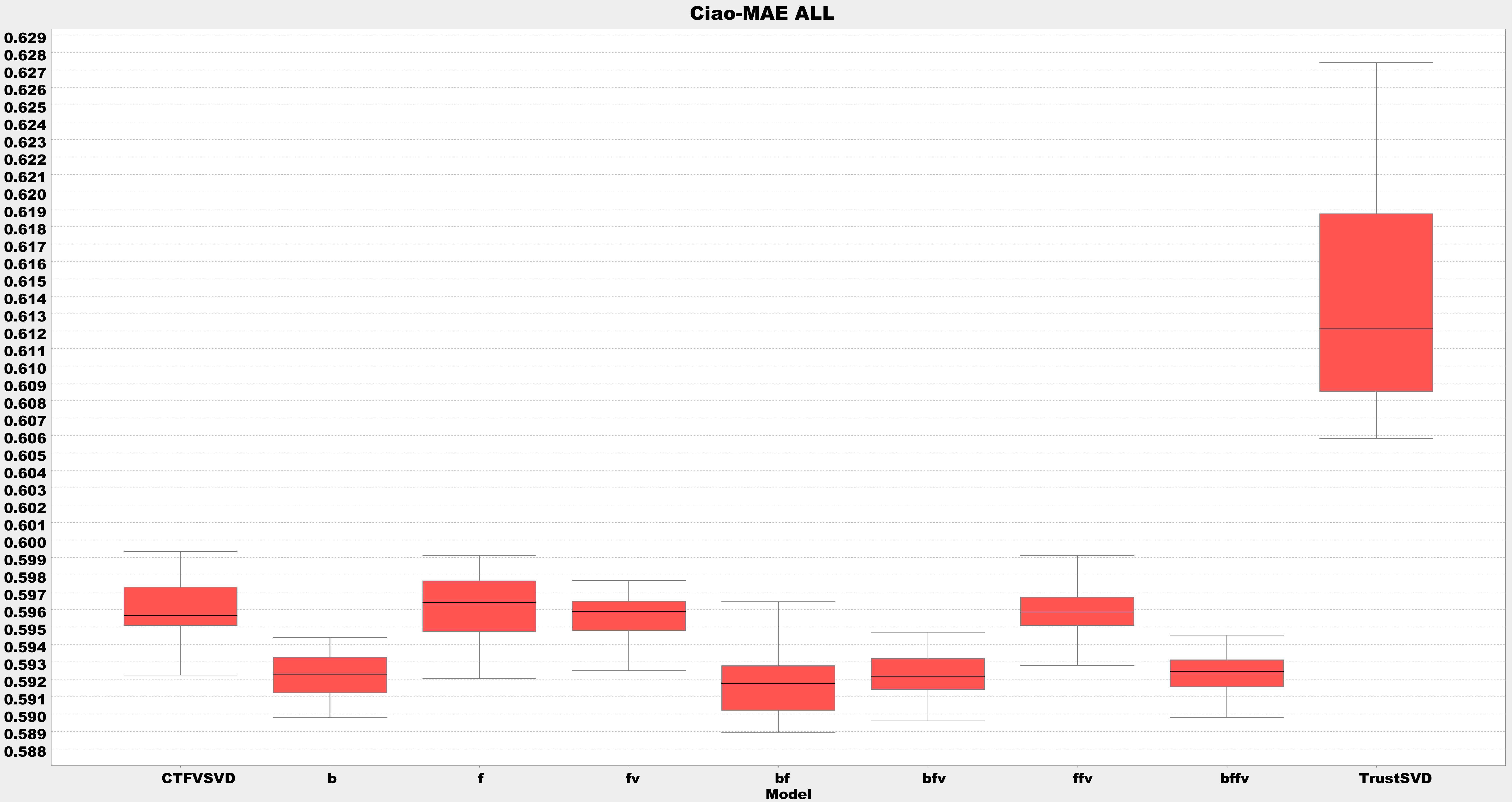}
		\caption{MAE, all users}
		\label{fig5a}
	\end{subfigure}%
	\begin{subfigure}[b]{0.5\textwidth}
		\includegraphics[height=5cm,width=6.5cm]{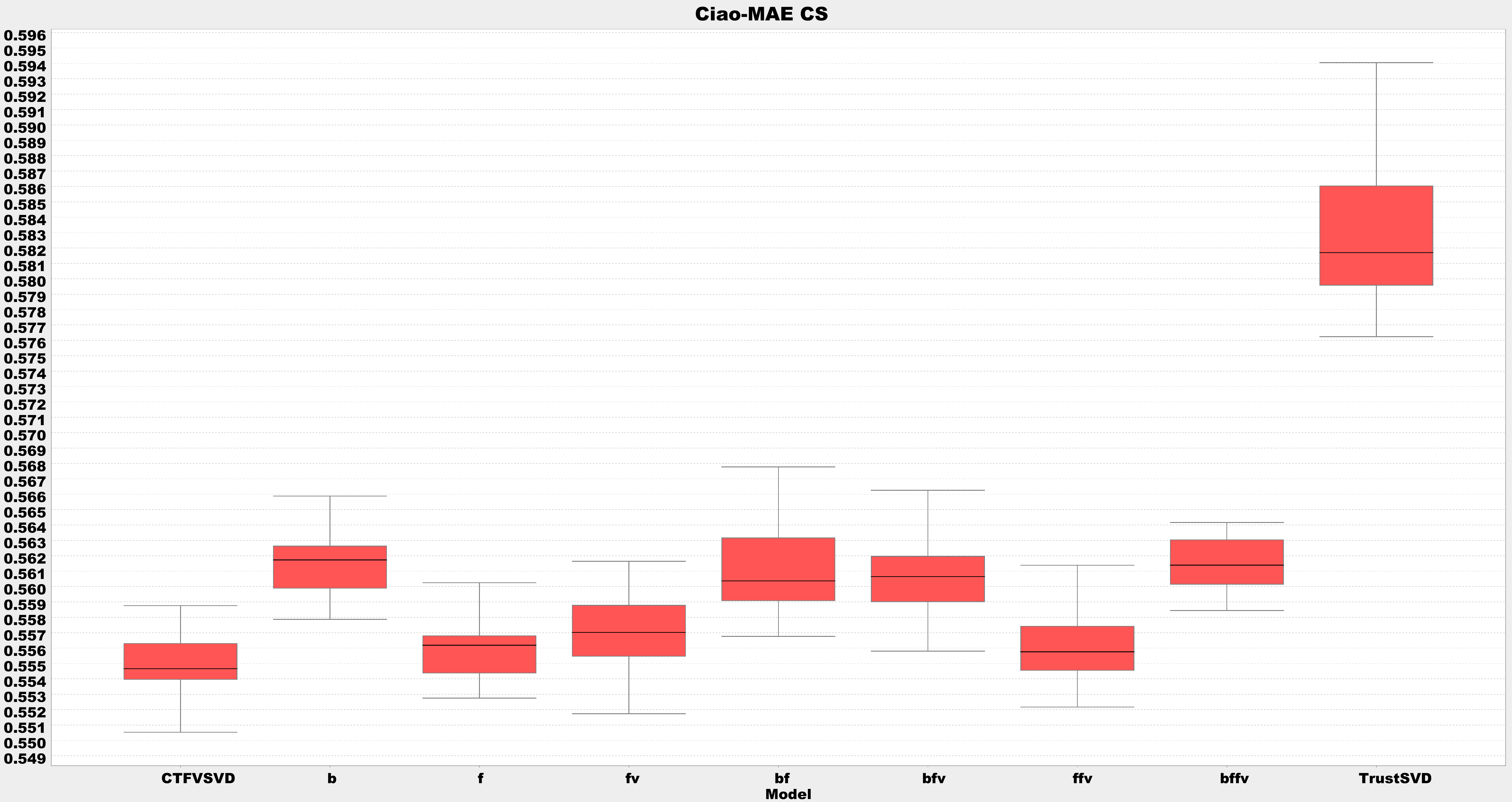}
		\caption{MAE, cold-start users}
		\label{fig5b}
	\end{subfigure}%
	\newline
	\begin{subfigure}[b]{0.5\textwidth}
		\includegraphics[height=5cm,width=6.5cm]{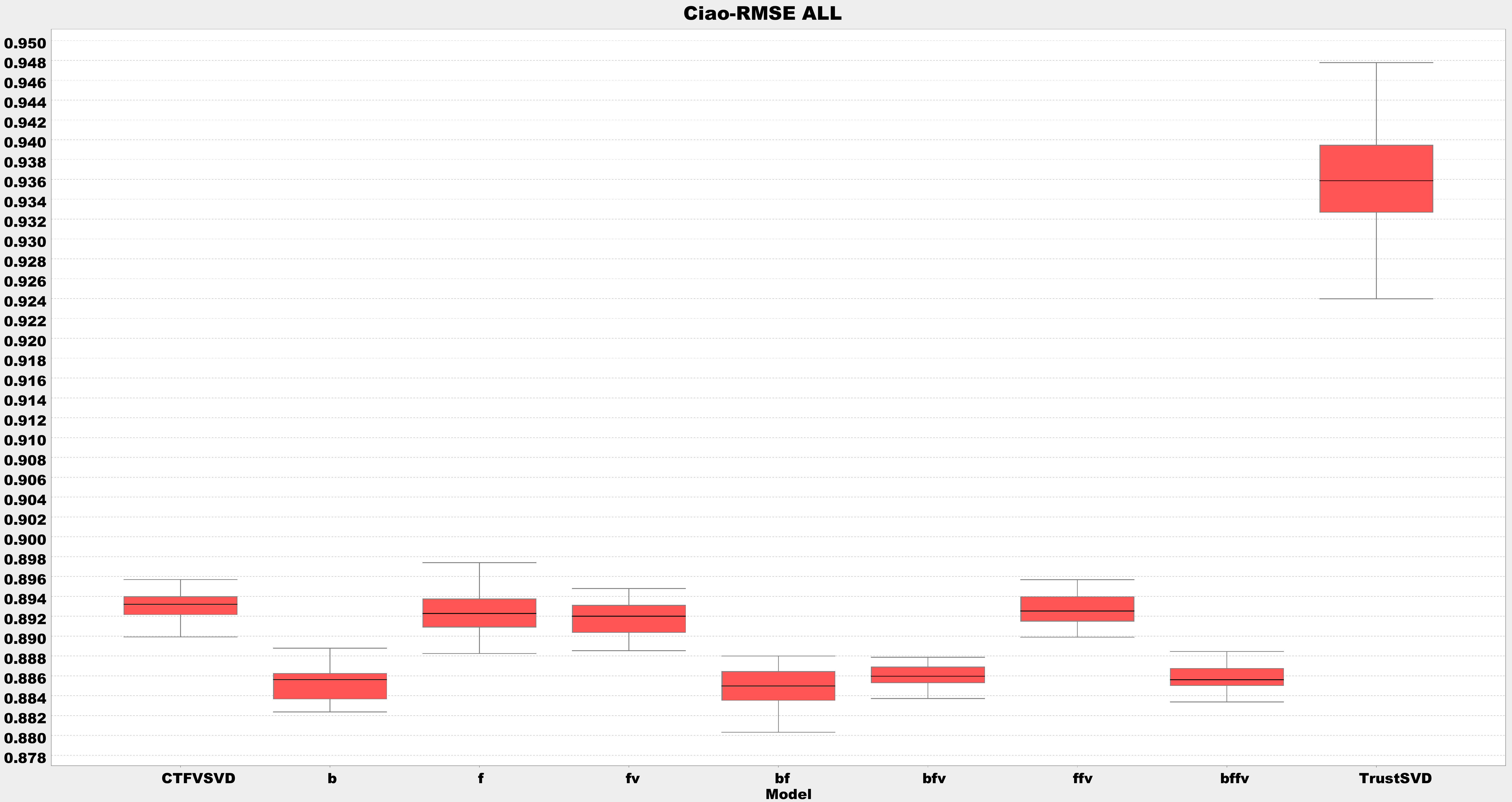}
		\caption{RMSE, all users}
		\label{fig5c}
	\end{subfigure}%
	\begin{subfigure}[b]{0.5\textwidth}
		\includegraphics[height=5cm,width=6.5cm]{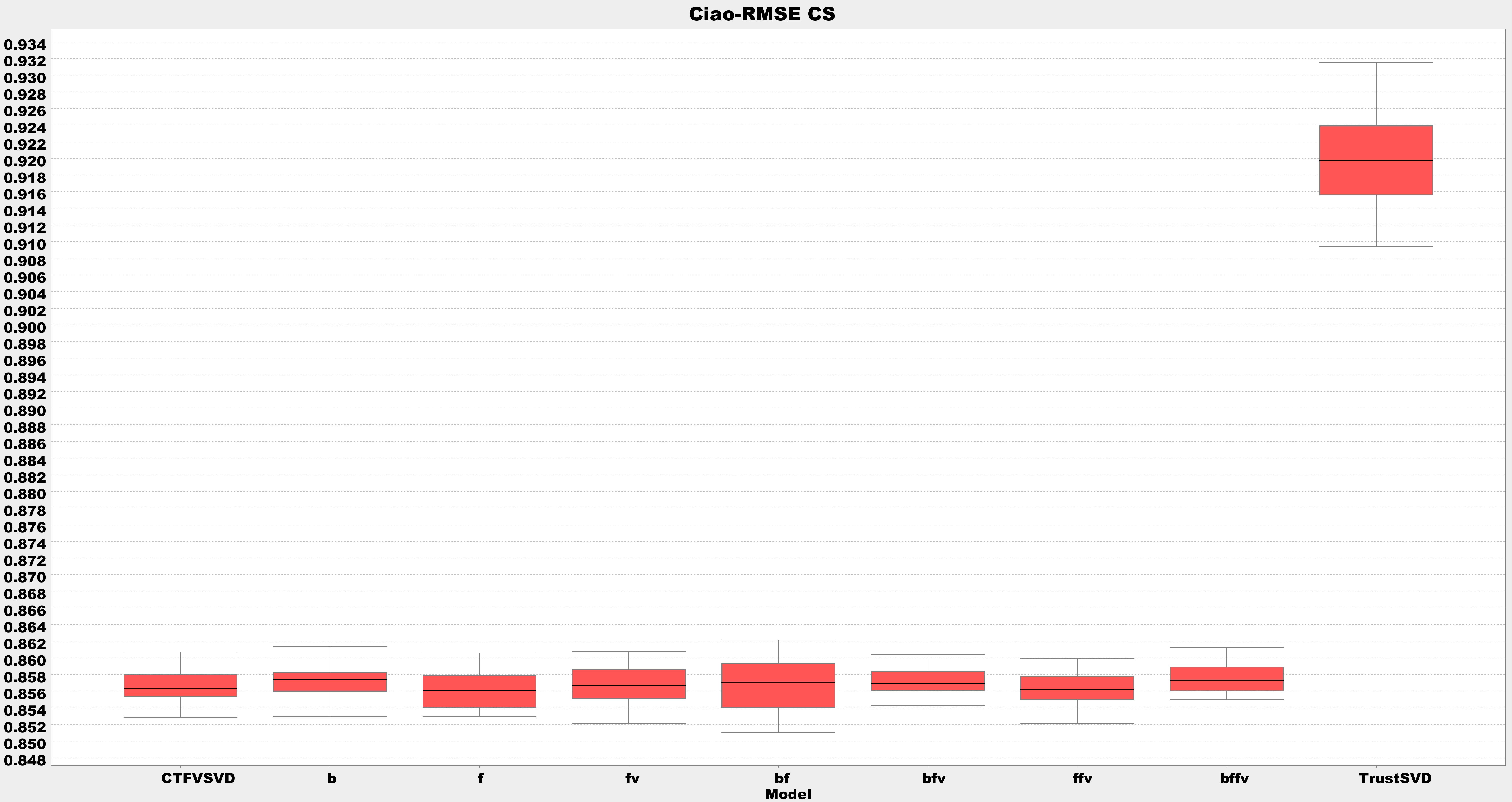}
		\caption{RMSE, cold-start users}
		\label{fig5d}
	\end{subfigure}%
	\centering
	\caption{Box plots of the Aspect-MF's combinations (b, bf, bffv, f, ffv, fv) and CTFVSVD versus TrustSVD in Ciao dataset in terms of MAE and RMSE measures for cold-start users (CS) and all users (ALL).}
	\label{fig5}
\end{figure}

\begin{figure}[!ht]
	\vskip 0cm
	\centering
	\setcounter{figure}{5}
	\begin{subfigure}[b]{0.5\textwidth}
		\includegraphics[height=5cm,width=6.5cm]{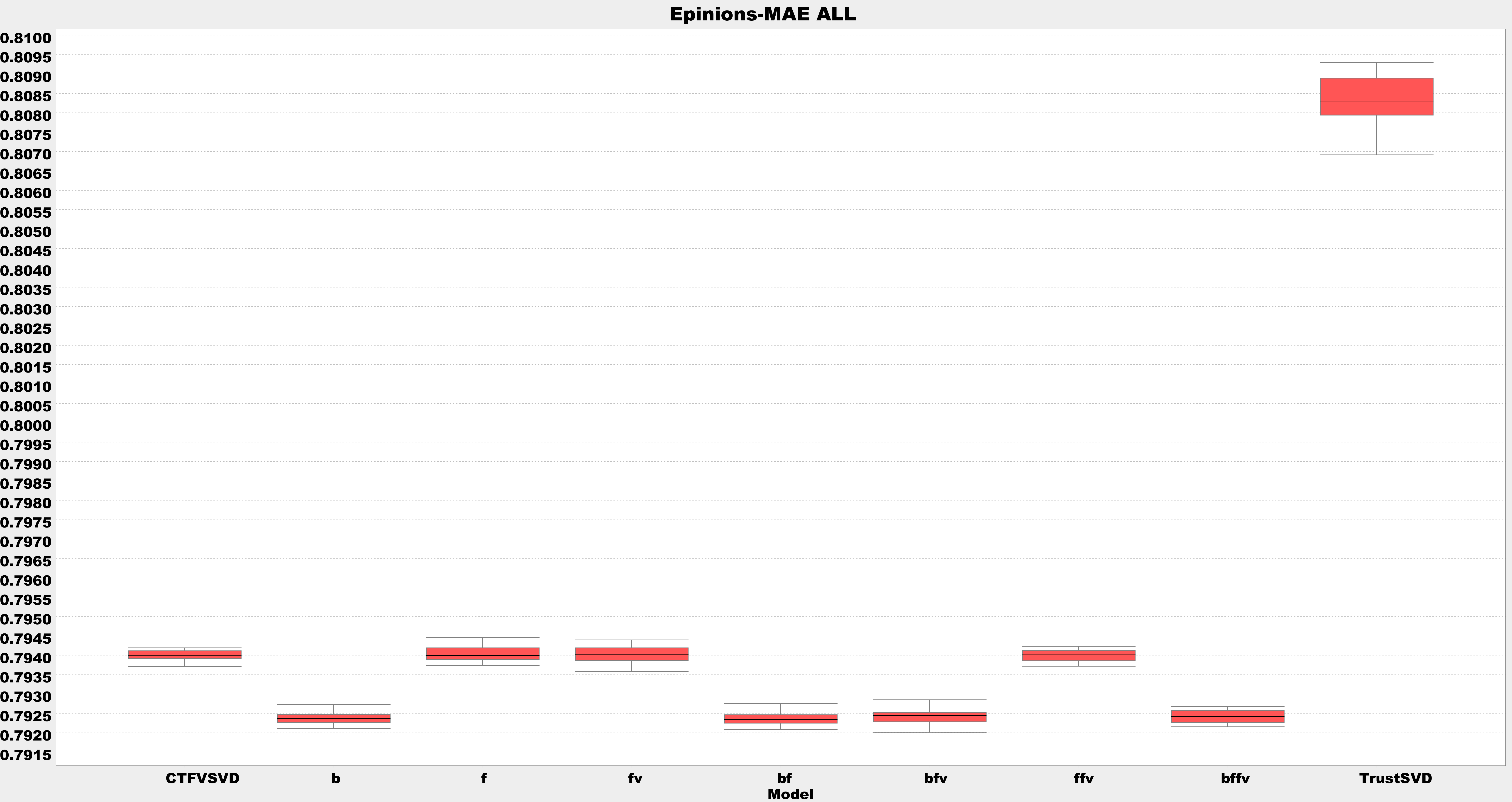}
		\caption{MAE, all users}
		\label{fig6a}
	\end{subfigure}%
	\begin{subfigure}[b]{0.5\textwidth}
		\includegraphics[height=5cm,width=6.5cm]{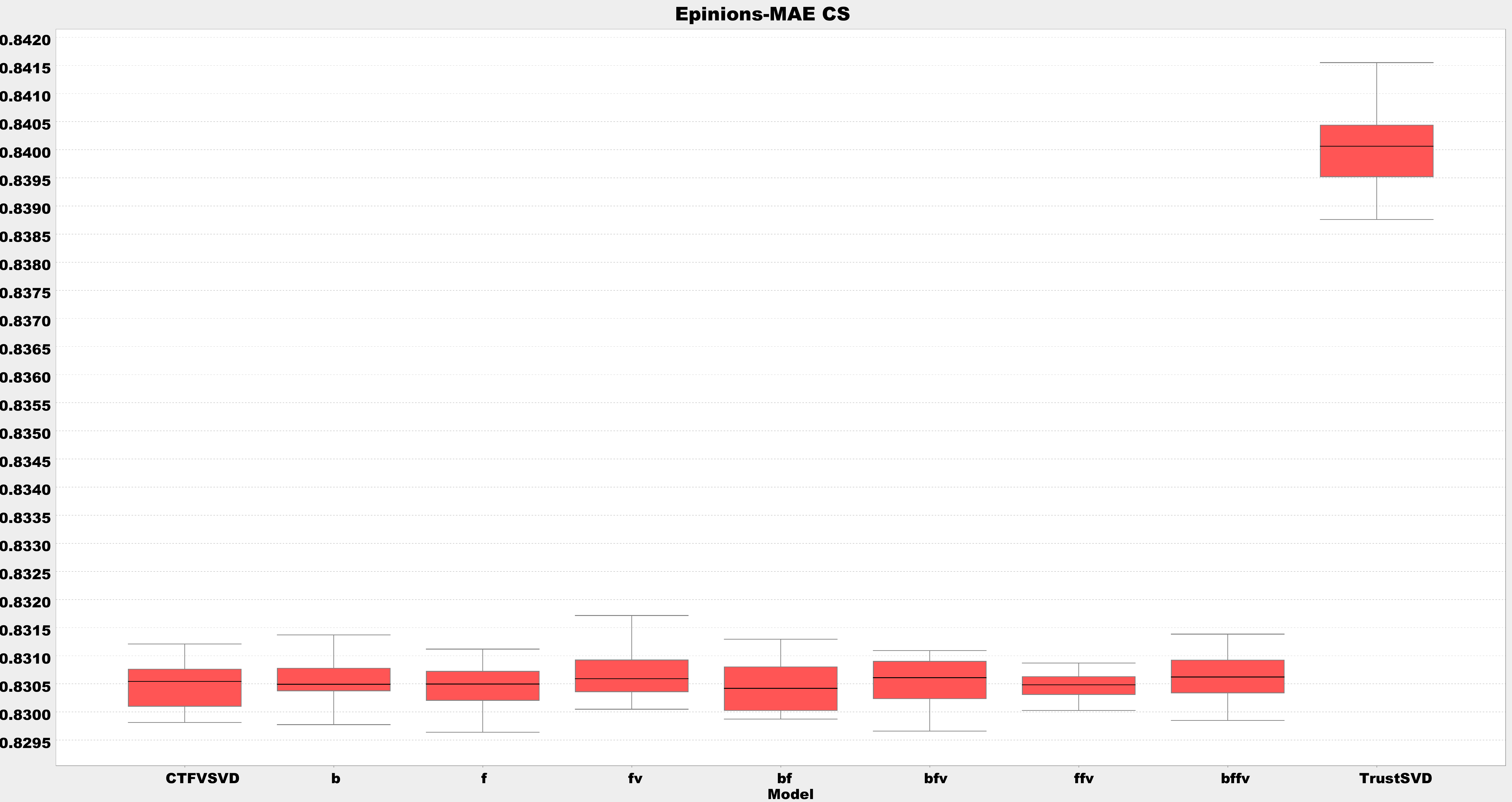}
		\caption{MAE, cold-start users}
		\label{fig6b}
	\end{subfigure}%
	\newline
	\begin{subfigure}[b]{0.5\textwidth}
		\includegraphics[height=5cm,width=6.5cm]{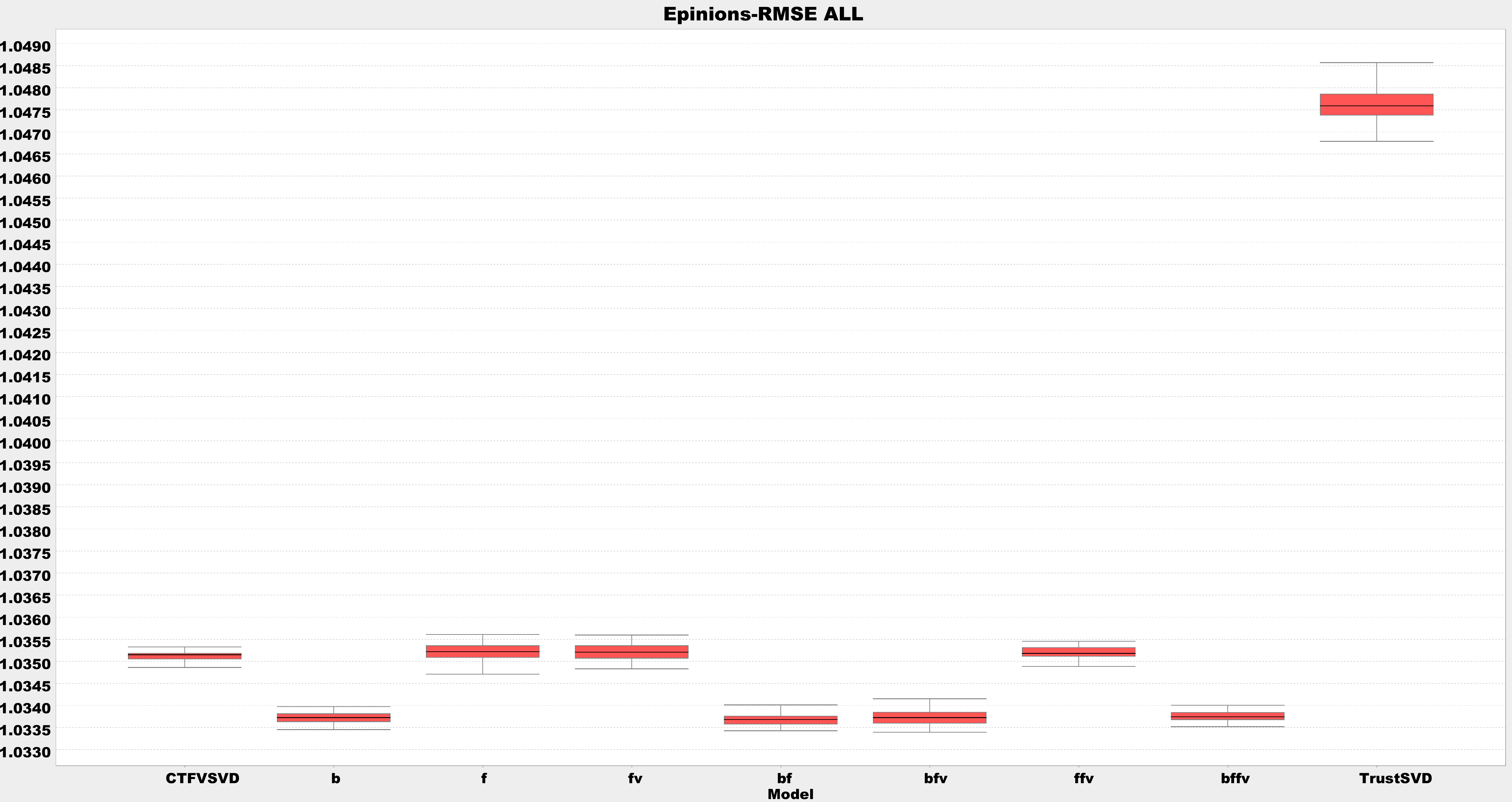}
		\caption{RMSE, all users}
		\label{fig6c}
	\end{subfigure}%
	\begin{subfigure}[b]{0.5\textwidth}
		\includegraphics[height=5cm,width=6.5cm]{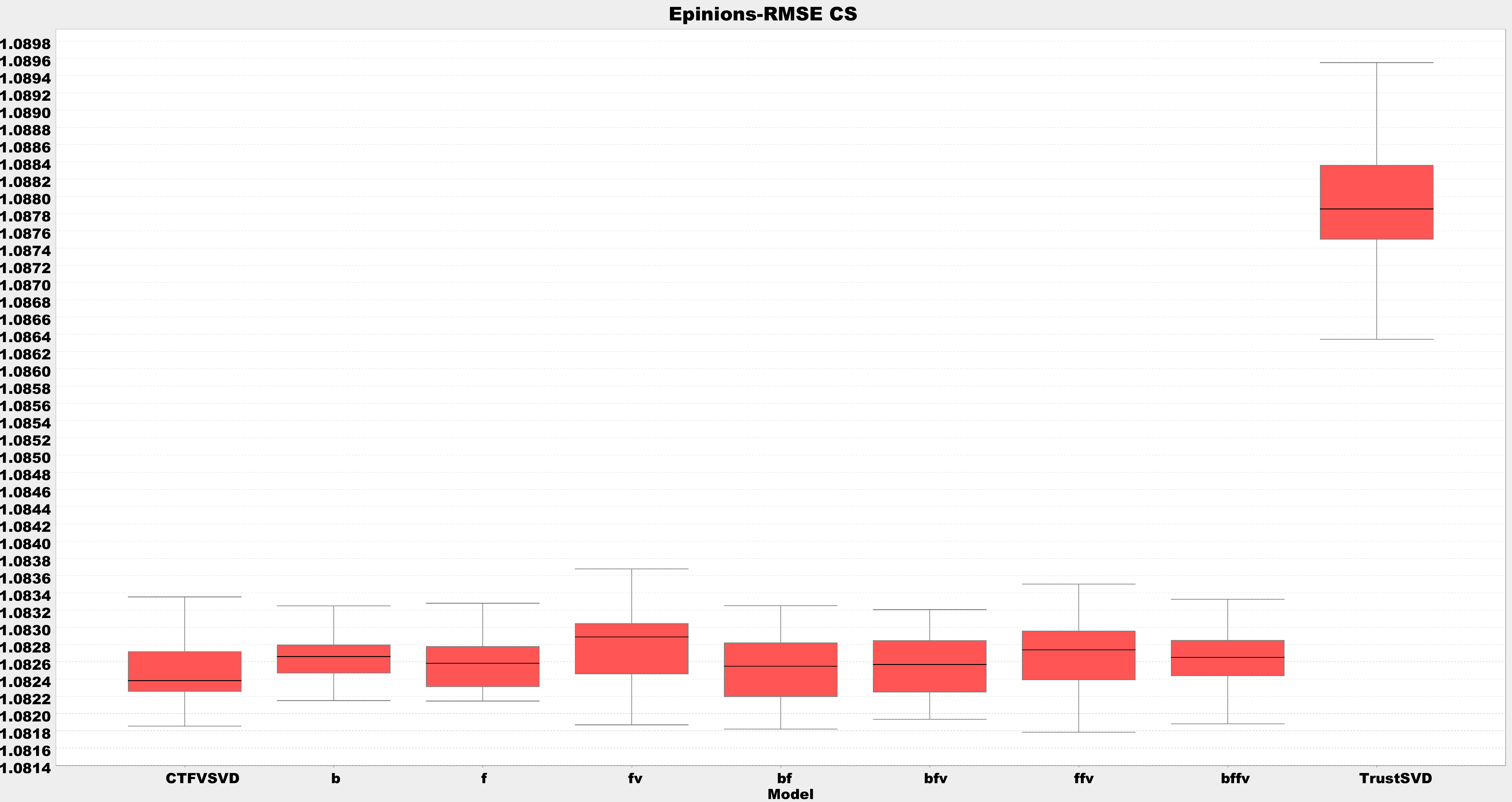}
		\caption{RMSE, cold-start users}
		\label{fig6d}
	\end{subfigure}%
	\centering
	\caption{Box plots of the Aspect-MF's combinations (b, bf, bffv, f, ffv, fv) and CTFVSVD versus TrustSVD in Epinions dataset in terms of MAE and RMSE measures for cold-start users (CS) and all users (ALL).}
	\label{fig6}
\end{figure}

\begin{figure}[!ht]
	\vskip 0cm
	\centering
	\setcounter{figure}{6}
	\begin{subfigure}[b]{0.5\textwidth}
		\includegraphics[height=5cm,width=6.5cm]{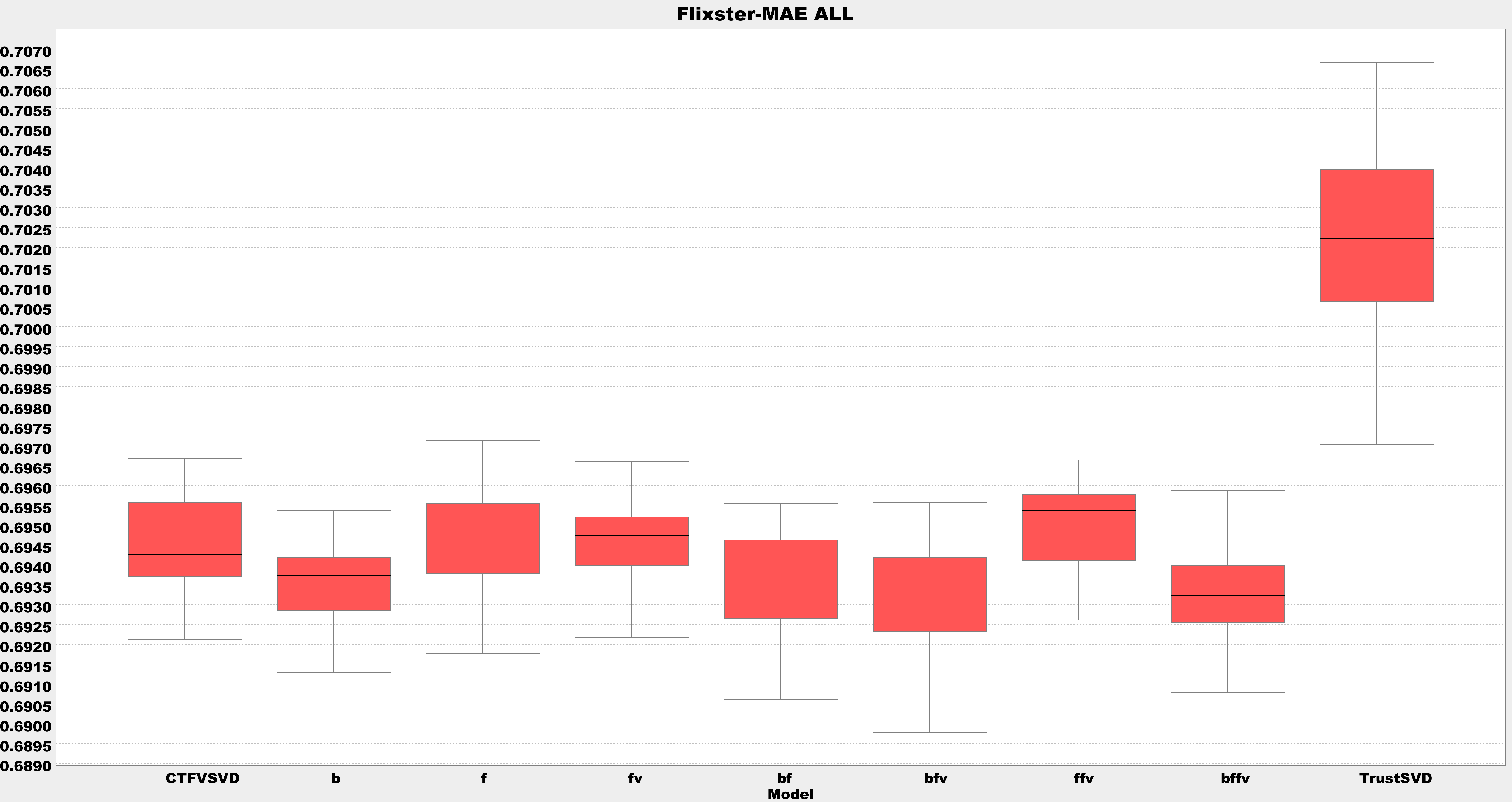}
		\caption{MAE, all users}
		\label{fig7a}
	\end{subfigure}%
	\begin{subfigure}[b]{0.5\textwidth}
		\includegraphics[height=5cm,width=6.5cm]{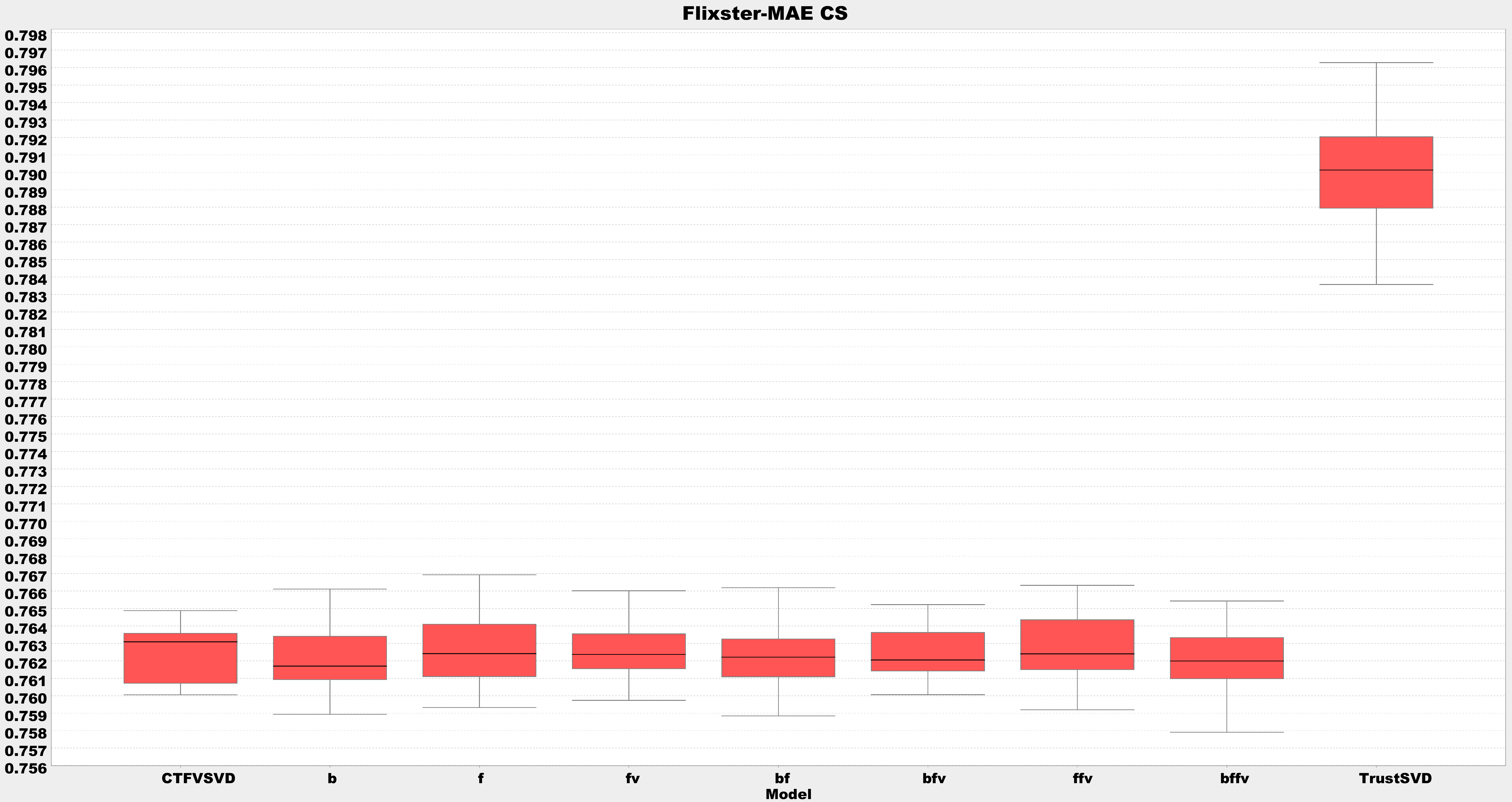}
		\caption{MAE, cold-start users}
		\label{fig7b}
	\end{subfigure}%
	\newline
	\begin{subfigure}[b]{0.5\textwidth}
		\includegraphics[height=5cm,width=6.5cm]{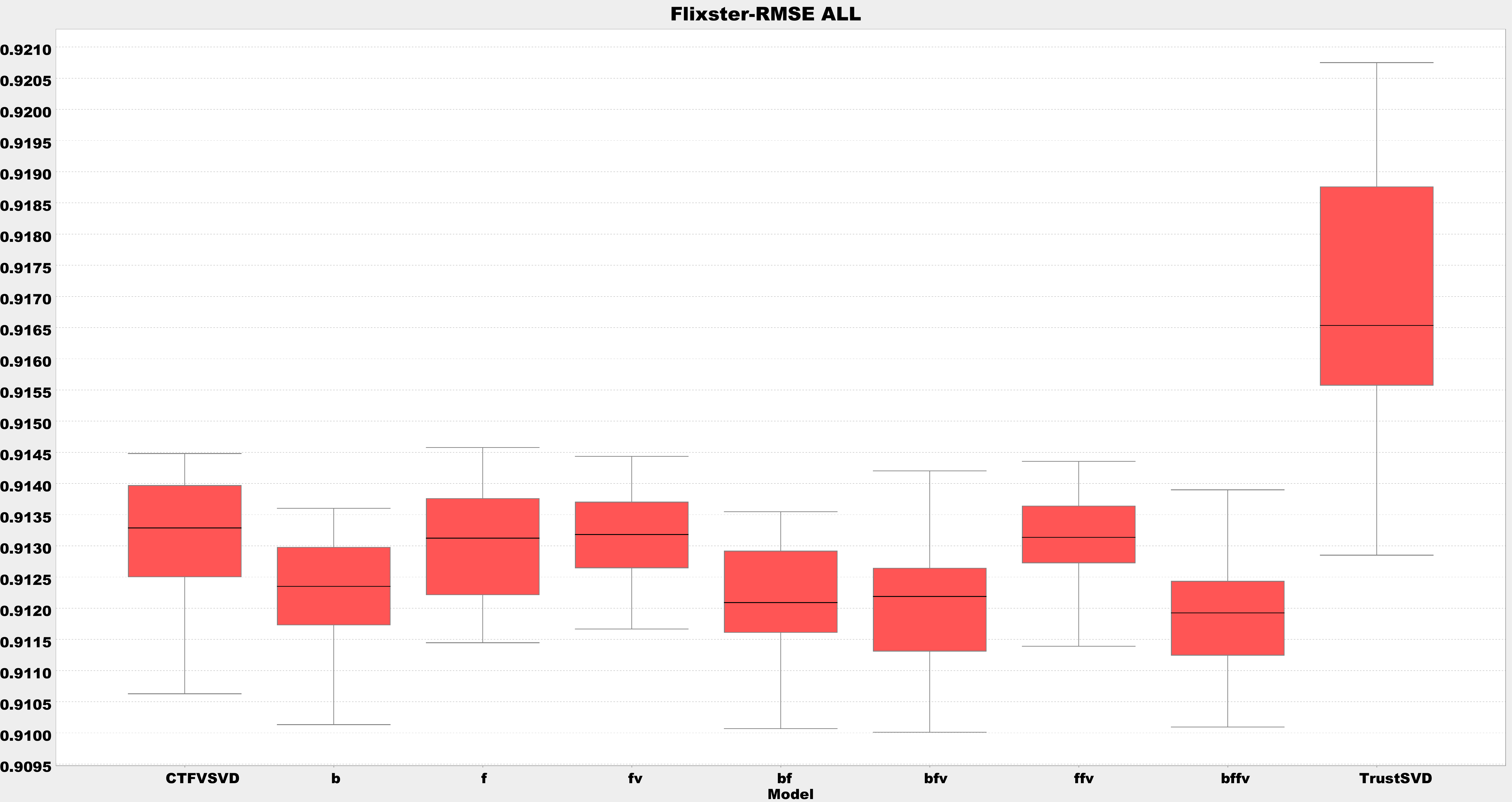}
		\caption{RMSE, all users}
		\label{fig7c}
	\end{subfigure}%
	\begin{subfigure}[b]{0.5\textwidth}
		\includegraphics[height=5cm,width=6.5cm]{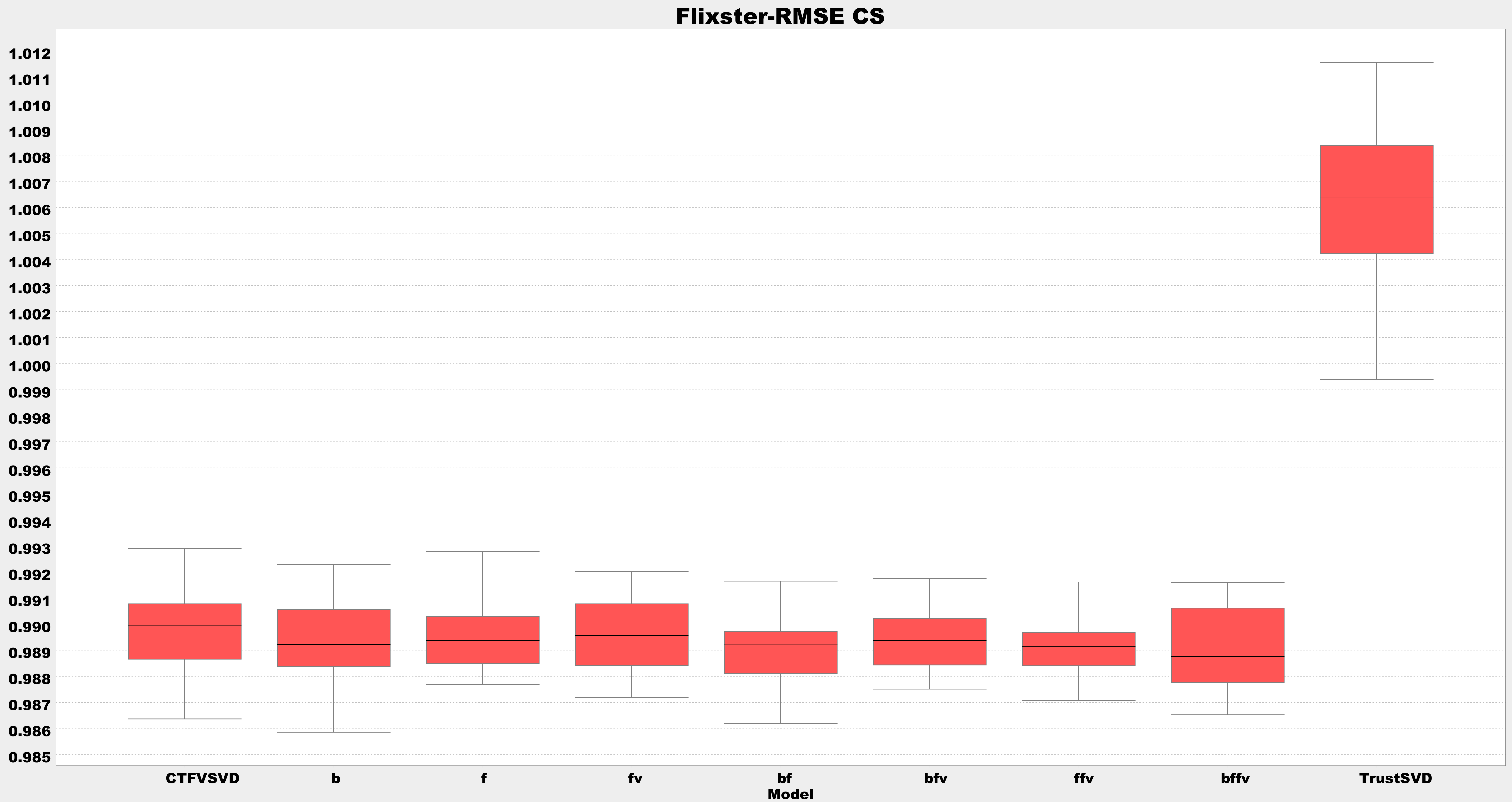}
		\caption{RMSE, cold-start users}
		\label{fig7d}
	\end{subfigure}%
	\centering
	\caption{Box plots of the Aspect-MF's combinations (b, bf, bffv, f, ffv, fv) and CTFVSVD versus TrustSVD in Flixster dataset in terms of MAE and RMSE measures for cold-start users (CS) and all users (ALL).}
	\label{fig7}
\end{figure}

\paragraph{\textbf{Accuracy measures}}
\label{Accuracy Measures}
As the statistical analysis of the models in Table \ref{table1} show, the differences are generally more significant when the accuracies are measured in terms of the RMSE. This can be explained by the formulation of these models as an optimisation problem. These models focus on maximising accuracy using RMSE and achieving better MAE values is a secondary goal that is only pursued through minimising RMSE. 

\paragraph{\textbf{Cold-start vs all users}}
\label{Cold-start vs All Users}
By taking a close look at the statistical analysis results in Table \ref{table1} and also the box plots of CTFVSVD vs Aspect-MF's combinations in Fig. \ref{fig5}, we can see that in all three datasets, the improvements of the Aspect-MF are more significant over all users than cold-start users. This can be explained by the amount of dynamic information that the models receive for each one of these groups of users. For all users, the model is trained using all ratings and also all associated time stamps for those ratings. Therefore the model can more successfully discern the temporal patterns in the preferences, and the accuracy improvements are larger. However, for the cold-start users, the model does not have access to much temporal information about these users, since they do not have many ratings. As a result, the model cannot identify the shift in the preferences of these users, and the improvements are smaller. From this, we conclude that temporal models are more successful on all users, because for them, temporal information is available.

\subsection{Statistical analysis}
\label{Statistical Analysis}
The statistical analysis of the performances provided in Table \ref{table1} shows that all Aspect-MF's combinations achieve significantly better results than TrustSVD, which does not include the temporal information. The values in Table \ref{table2} also show that Aspect-MF's combinations also result in improvements over CTFVSVD that are statistically significant, which means that in all three datasets, Aspect-MF has been successful in extracting the temporal patterns in the users' preferences. We can also see that the all the p values in Table \ref{table1} are 0.0000, which means that with almost 100\% probability, the two model executions (Aspect-MF and TrustSVD) do not come from distributions with equal mean performances. Therefore, we are almost 100\% sure that the observed differences in performance are due to the superiority of Aspect-MF over TrustSVD, and not the result of chance. Similarly, the p values in Table \ref{table2} are almost zero, which means that we are certain that Aspect-MF is better than CTFVSVD, in cases where the t test shows a statistically significant improvement.

\subsection{Dynamic aspects}
\label{Dynamic aspects}
The close comparison of the error values achieved by Aspect-MF in Fig. \ref{fig8} show that in terms of MAE for all users, Aspect-MF achieves the best performance on the Ciao and Epinions datasets, for the models including dynamic $b$ and $f$ aspects. However, on the Flixster dataset, the model combination with dynamic $b$ and $fv$ aspects performs best. Interestingly, for cold-start users, different models perform the best. In particular, on the Ciao dataset, the model including dynamic $f$ performs best, whereas on the Epinions and Flixster datasets, the model including dynamic $b$, $f$, and $fv$ aspects, and the model with drifting $f$ aspect achieve the best results respectively. Furthermore, the error values in Fig. \ref{fig9} show that different model combinations might achieve the best performances for RMSE.
From these figures, we can make several conclusions.

The first conclusion is that the dynamic patterns are dataset-dependent. Therefore, users and the items in different dataset can have preferences with aspects with different levels of dynamicity. This finding supports our component-based approach in modelling the dynamic properties of the preference aspects.

The second conclusion is that the prediction of the ratings for the cold-start users is less dependent on the drifting bias than that of all users. As we see in this Figures \ref{fig8} and \ref{fig9}, for all users, the combinations that include dynamic $b$ aspects are strictly better than the other combinations, whilst this is less consistent for cold-start users, where sometimes the models with only dynamic $f$ aspects perform best. This suggests that the preferences of cold-start users are not much affected by the shifts in the popularity of the items, while other users' preferences are more influenced by such shifts. Therefore, the accurate modelling of such temporal effects is of greater importance in the case of all users than cold-start users. As previous studies have shown \cite{koenigstein2011yahoo}, bias is a very important aspect in human preferences. Since the cold-start users do not have enough ratings, there is also not enough temporal data to train the preferences for these models. Therefore, the trained temporal aspects of these users are probably not very accurate, and therefore, the combinations that include bias perform poorly on these users, due to imprecise predictions.

The third conclusion is that both measures reveal roughly the same preference patterns. This seems justifiable, since the shift in user preferences should naturally be independent of how the differences in estimated preferences and real preferences are measured.

To summarise, it is very advantageous to have a component-based model in which the temporal aspects of preferences can be arbitrarily captured in different conditions. This enables us to capture the patterns only when they are actually helpful, and consequently, build the most accurate preference models, tailored to different datasets and domains with disparate temporal patterns.

\begin{figure}[H]
	\setcounter{figure}{0}
	\includegraphics[width=1\linewidth]{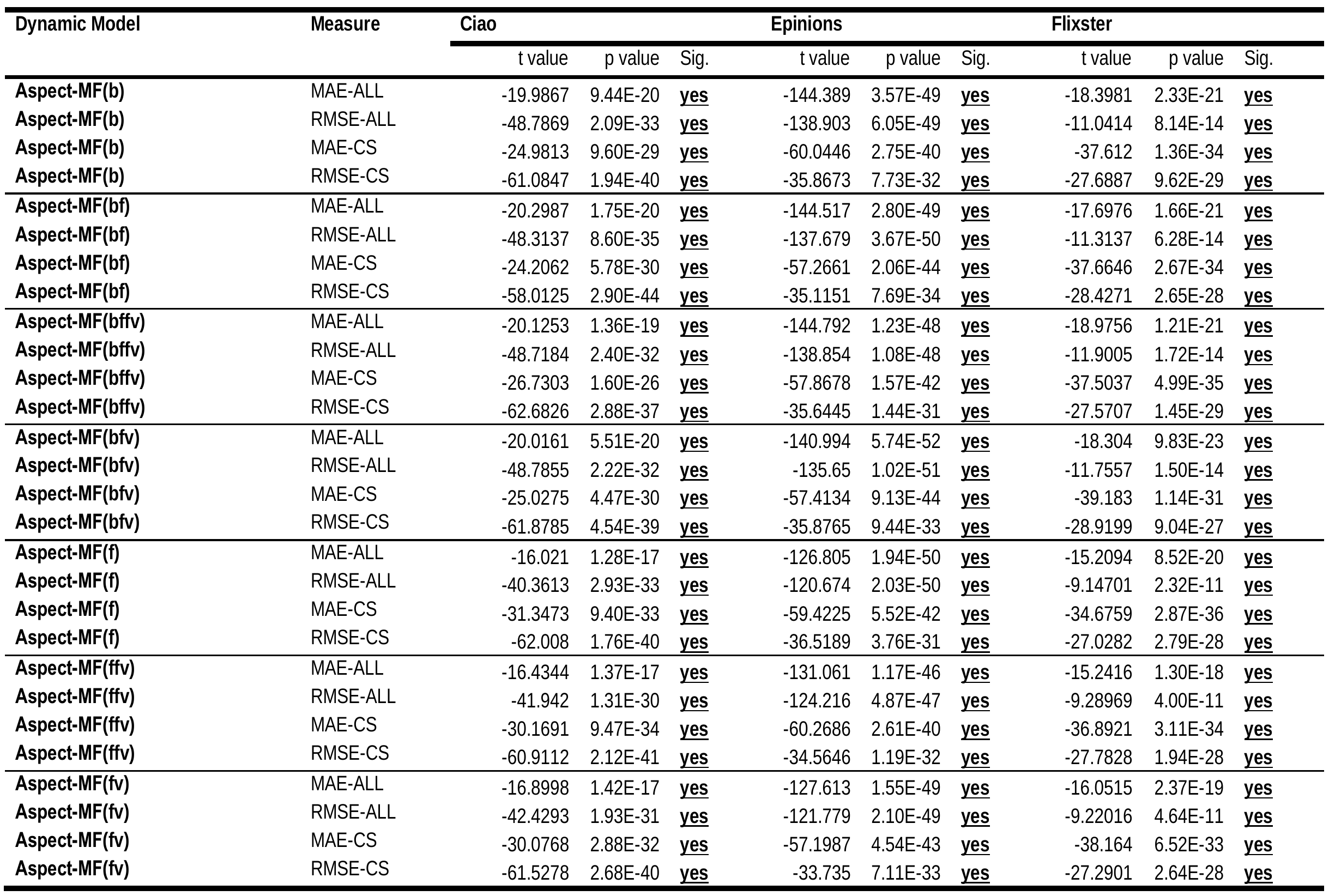}
	\captionsetup{labelformat=empty}
	\centering
	\caption{Table 1: The t values and p values for Aspect-MF's combinations vs TrustSVD in Ciao, Epinions, and Flixster datasets for MAE and RMSE measures on all users (ALL) and cold-start users (CS)}
	\label{table1}
\end{figure}

\begin{figure}[H]
	\setcounter{figure}{1}
	\includegraphics[width=1\linewidth]{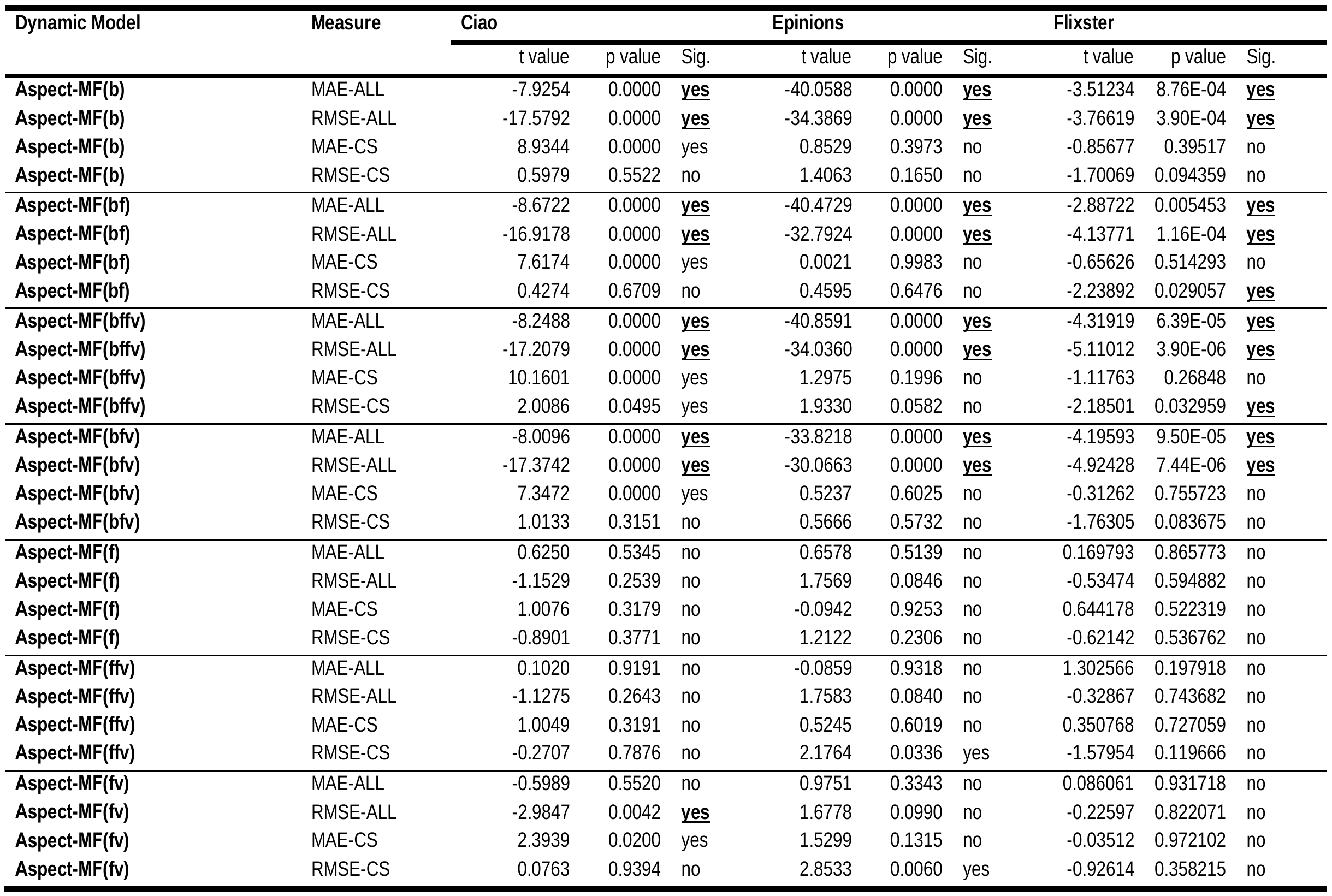}
	\captionsetup{labelformat=empty}
	\centering
	\caption{Table 2: The t values and p values for Aspect-MF's combinations vs CTFVSVD in Ciao, Epinions, and Flixster datasets for MAE and RMSE measures on all users (ALL) and cold-start users (CS)}
	\label{table2}
\end{figure}

\begin{figure}[!htp]
	\hspace*{0.2in}
	\centerline{\includegraphics[scale=0.7]{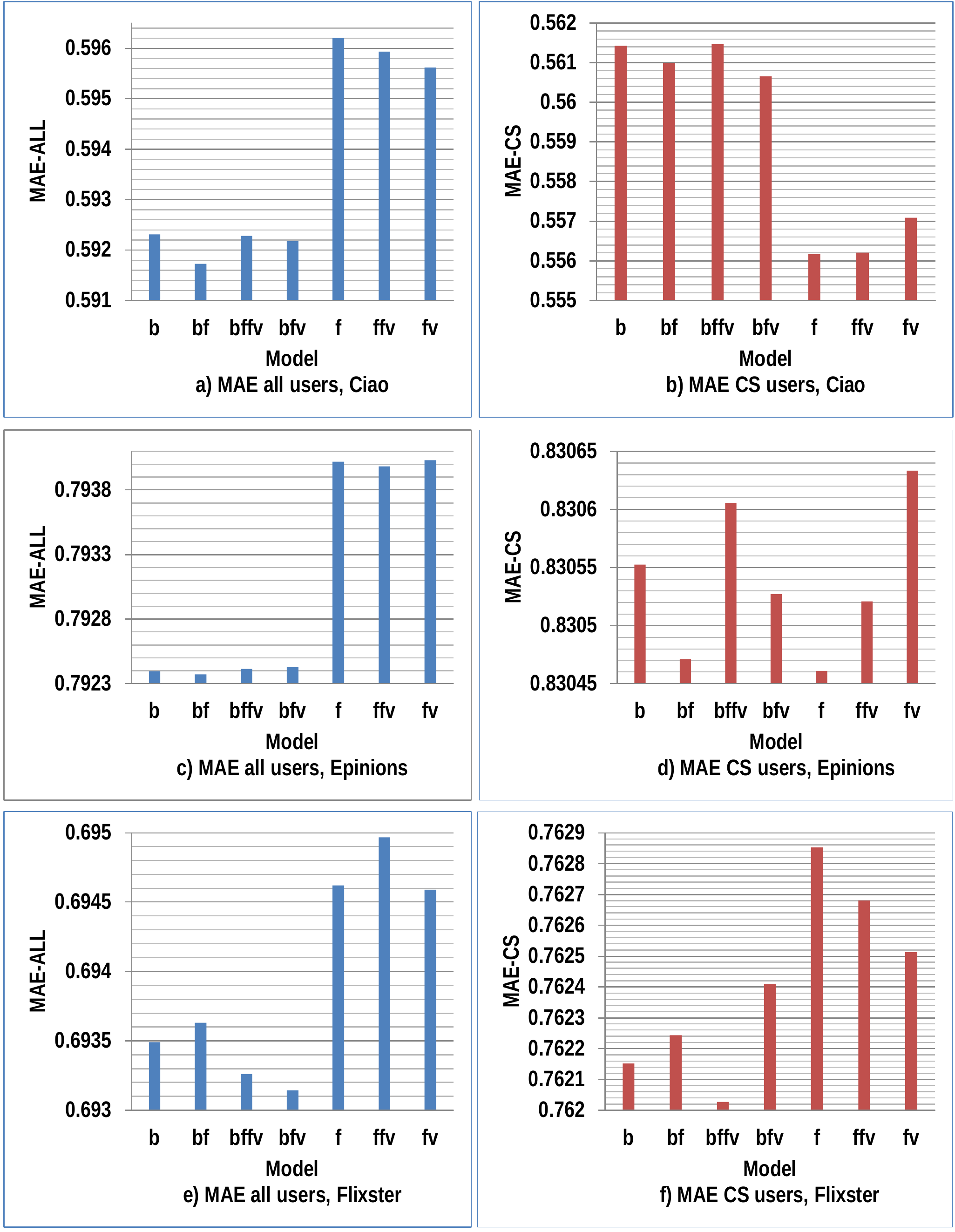}}
	\caption{Comparisons of the MAE values of Aspect-MF's combinations in a,b) Ciao, c,d) Epinions, and e,f) Flixster datasets for all users (MAE-ALL) and cold-start users (MAE-CS)}
	\label{fig8}
\end{figure}

\begin{figure}[!htp]
	\hspace*{0.2in}
	\centerline{\includegraphics[scale=0.7]{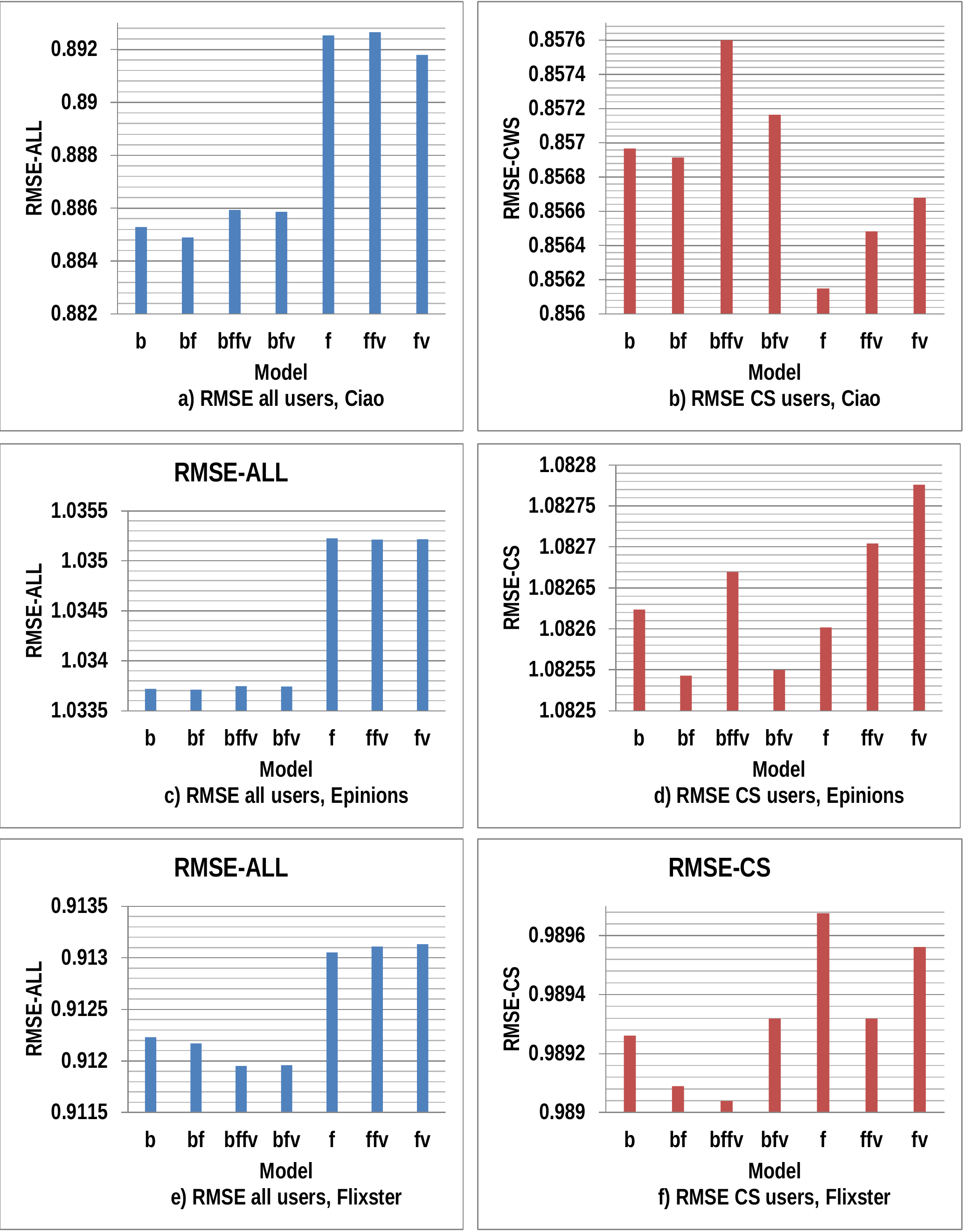}}
	\caption{Comparisons of the MAE values of Aspect-MF's combinations in a,b) Ciao, c,d) Epinions, and e,f) Flixster datasets for all users (RMSE-ALL) and cold-start users (RMSE-CS)}
	\label{fig9}
\end{figure}

\subsection{Effect of the size of the training dataset}
\label{Effect of the size of the training dataset}
The main purpose of this section is to evaluate the robustness of the models against shortage of training data. In the experiments in sections \ref{Discussion} through \ref{Dynamic aspects}, 80\% of the ratings matrix was used for training the models and the remaining data was used for evaluation. The question that arises here is how the models would perform if less amount of data was fed to the models for training.

\begin{figure}[!htp]
	\vskip 0cm
	\centerline{\includegraphics[scale=0.3]{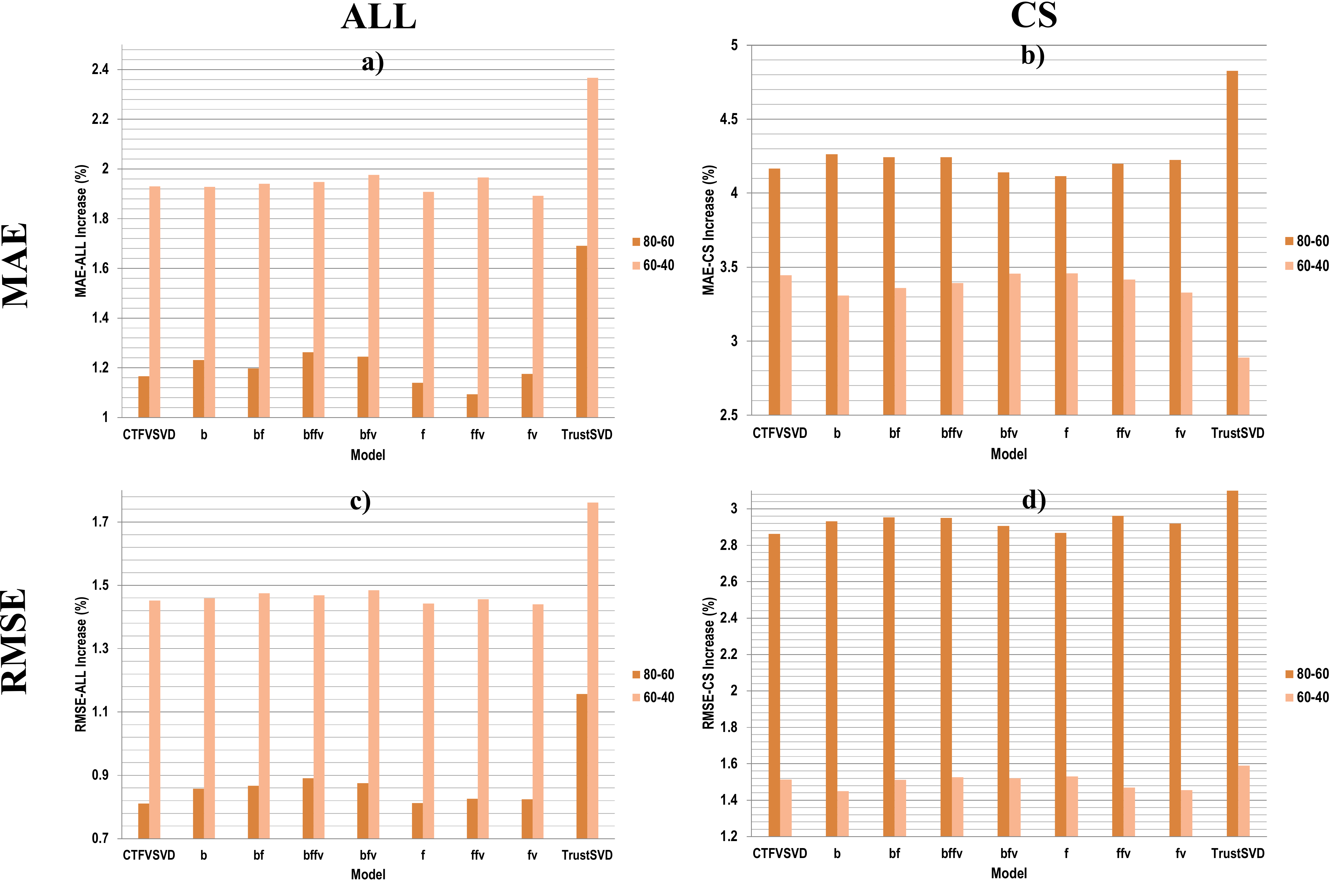}}
	\caption{Effect of the training amount on Flixster dataset for a) MAE for all users, b) MAE for cold-start users, c) RMSE for all users, d) RMSE for cold-start users}
	\label{fig10}
\end{figure}

\begin{figure}[!htp]
	\vskip 0cm
	\centerline{\includegraphics[scale=0.3]{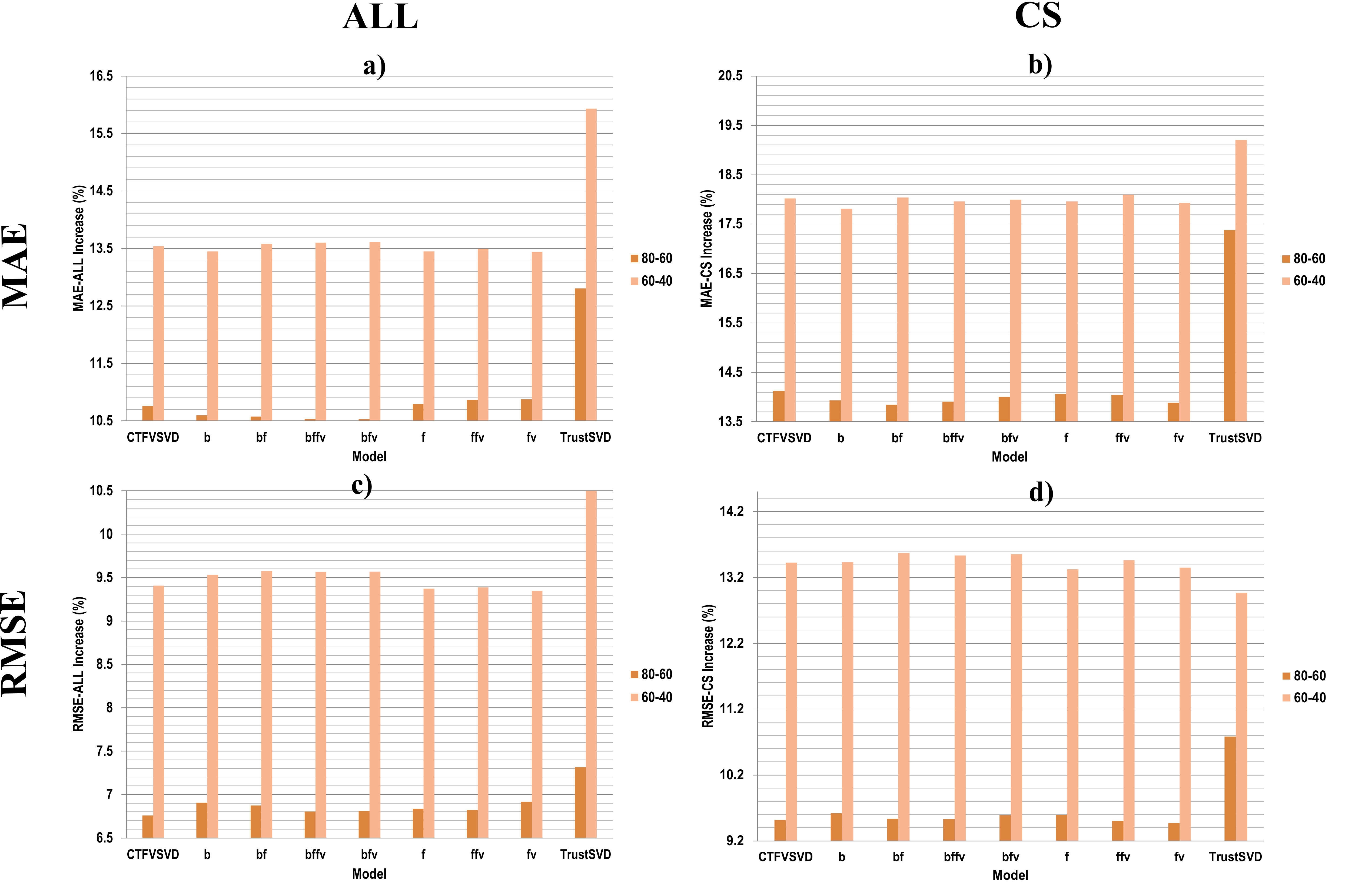}}
	\caption{Fig. 11: Effect of the training amount on Ciao dataset for a) MAE for all users, b) MAE for cold-start users, c) RMSE for all users, d) RMSE for cold-start users}
	\label{fig11}
\end{figure}

In order to analyse the behaviour of the models with respect to the amount of training data, we can reduce the amount of the training data, and consider how much the accuracy drops as the training data is decreased. Therefore, we also evaluate the models in two additional cases. The first case includes 60\% of the data for training, and the remaining 40\% for testing, and the second case uses 40\% of ratings data for training and the rest for evaluation. The results for the Flixster and Ciao datasets are demonstrated in Figs. \ref{fig10} and \ref{fig11} respectively. These figures show the percentage of error increase as the amount of training data is decreased.

\paragraph{\textbf{All users}}
As can be seen in Fig. \ref{fig10}, on the Flixster dataset, in the case of all users, all combinations of Aspect-MF result in a smaller increase in the error when the training data is decreased from 80\% to 60\% (denoted by 80-60 in these diagrams), and from 60\% to 40\% (denoted by 60-40 in these diagrams). Furthermore, we can observe that in terms of MAE, the combination that includes $f$ and $fv$ resulted the smallest error increase when the training data decreased from 80\% to 60\%, and the model that included $fv$ resulted in the smallest error increase when the training data decreased from 60\% to 40\%. This suggests that the dynamic model is more robust to the shortage of training data, when the error is measured in terms of MAE for all users. In terms of RMSE, the least accuracy deterioration happened for the model combination with the $f$ aspect, both when the training data amount drops to 60\%, and when it drops to 40\%.

\paragraph{\textbf{Cold-start users}}
For cold-start users however, a different pattern is evident. Interestingly, we can see that for cold-start users, the error increases more when the training data is decreased from 80\% to 60\%, compared to when it is decreased from 60\% to 40\%. This means that the accuracy degrades more when the training data drops to 60\%. Judging by the higher error increase for cold-start users in comparison with all users, cold-start users seem to be more sensitive to the decrease in the amount of training data. This seems understandable, since the cold-start users do not have many ratings. Therefore, when evaluating the model accuracy for cold-start users, less accurate predictions for each rating have a larger effect on the overall accuracy.

TrustSVD seems to be more robust to the shortage of training data for cold-start users, when the training data drops from 60\% to 40\%. This can be attributed to the fact that the dynamic model contains time information, and this information can be misleading if we substantially decrease the amount of training data, and evaluate the accuracy for cold-start users who do not have much ratings. A similar observation was made in Figs. \ref{fig8} and \ref{fig9}, where the dynamic model including the $b$ aspect performed poorly on the cold-start users.

\paragraph{\textbf{All users vs cold-start users}}
A similar trend to the one observed in Flixster dataset can also be seen in the Ciao dataset in Figure \ref{fig11}. As this figure shows, the accuracy deterioration for cold-start users is much larger compared with that for all users. Again, we attribute this to the high sensitivity of cold-start users to inaccurate predictions.
For the case where the training data amount drops from 80\% to 60\%, the model combination with all the dynamic aspects ($bffv$) results in the lowest increase in MAE for all users. For cold-start users, the model combination with $b$ and $f$ aspects achieve the smallest deterioration of accuracy.

However, in terms of RMSE for all users, TrustSVD incurs the lowest increase in the error, while for cold-start users, the model with the dynamic $fv$ aspect is the most robust. In the second case where the training data amount is decreased from 60\% to 40\%, at least one of the model combinations performs best (incurs the lowest accuracy deterioration) for each measure, among the models tested. We can also see that when the training data amount is decreased from 80\% to 60\%, the error increase is much lower than when the training data amount drops from 60\% to 40\%. This means that the models are still quite robust with 60\% of the ratings data as training data, but their accuracy considerably drops when the training data decreases to 40\%.

\paragraph{\textbf{Flixster vs Ciao}}
One of the key differences between the behaviour of the models on the Flixster and Ciao datasets, as can be seen in Figs. \ref{fig10} and \ref{fig11}, is the threshold at which the accuracy sharply drops for cold-start users. For the Flixster dataset, the accuracy of cold-start users sharply worsens when the training data amount is decreased from 80\% to 60\%, while for the Ciao dataset, the sharp decrease in accuracy happens when the training data amount decreases from 60\% to 40\%. This can be easily justified by looking at the statistics of these two datasets for cold-start users. On the Flixster dataset as we mentioned before, each cold-start user rates 1.94 items on average, while this number is 2.94 in the Ciao dataset. Therefore, the accuracy of cold-start users on the Flixster dataset is more sensitive to inaccurate predictions than that on the Ciao dataset.

Considering all four measures on the two datasets, in general, we can observe that Aspect-MF's combinations are more robust to the decrease in the amount of training information than TrustSVD and CTFVSVD. The combinations  in this paper are particularly more helpful in cases where enough time related data is fed into the model as input.

\paragraph{\textbf{Insights}}
From the observations for cold-start users, we can conclude that in order for the time information to be helpful, we need to provide the model with enough time-related data as input, so that the accuracy can be improved, and the importance of such data is more pronounced for the cold-start users, whose predictions are more sensitive to the inaccuracies. Otherwise, if the amount of training data is insufficient, the model can learn unrealistic temporal patterns that directly result from a shortage of training information.

We also saw that the degree of deterioration of the accuracy is somewhat dependent on the dataset. On the Flixster, the accuracy degrades somewhere between just under 1\% to just under 5\%. On Ciao, however, the accuracy deteriorates much more (roughly between 6.5\% and 19.5\%). Therefore, it is up to the system users to decide whether they would like to use smaller datasets and sacrifice the accuracy, or spend more time on training more accurate models using more information. We did not observe any tangible differences between the execution times of these cases (80\%-60\%-40\%), and the computational complexity analysis of the model in section \ref{Computational Complexity Analysis} showed that the model time is of linear order. Therefore, it is probably advisable for the system owners to use as much data as available to achieve the highest accuracies, as long as their computational limitations allow.

\section{Conclusion and future work}
\label{Conclusion and Future Work}

In this paper, we addressed the problem of modelling the temporal properties of human preferences in recommender systems. In order to tackle this problem, we proposed a novel latent factor model called Aspect-MF. Aspect-MF built on the basis of CTFVSVD, a model that we proposed earlier, in order to capture socially-influenced conditional preferences over feature values. In Aspect-MF, three major preference aspects were assumed to be subject to temporal drift. These aspects included user and item biases, preferences over features, and preferences over feature values. Moreover, we also analysed the temporal behaviour of each of these preference aspects and their combinations. We also considered the robustness of Aspect-MF's combinations with respect to the shortage of training data.

In order to evaluate the model, we carried out extensive experiments on three popular datasets in the area of recommender systems. We considered the model errors in terms of MAE and RMSE measures on all users and cold-start users. We also performed statistical analyses on the performances observed, to make sure that the differences in accuracies are significant, and hence do not happen by chance. The experiments revealed that in all three datasets, all combinations of Aspect-MF for both measures on all users and cold-start users significantly outperformed TrustSVD, which had proven to be the most accurate static social recommendation model before CTFVSVD. The experiments also proved that most of the Aspect-MF's combinations were significantly more accurate than CTFVSVD. In particular, we found that Aspect-MF with all dynamic aspects outperformed CTFVSVD in all three datasets on all users.

The analysis of the temporal behaviour of preference aspects and their combinations on the three datasets showed that different datasets included different temporal patterns, and therefore, required models with different dynamic aspects. This supported our component-based approach in modelling the basic preference aspects and their temporal properties. We also concluded that the dynamic models are more helpful in cases there is enough training data to discern the temporal properties. In particular, we concluded that the models proposed in this paper are more successful in modelling all users, because more time-related data is available for all users than cold-start users, and therefore the temporal characteristics were extracted more accurately. The analysis of the robustness of the models with respect to the shortage of training data also revealed that Aspect-MF was in general more robust than CTFVSVD and TrustSVD. The models were also more robust for all users than cold-start users, because cold-start users were more sensitive to the inaccurate predictions.

A direction that we would like to pursue in the future is related to explaining the resulting recommendations to the users. Explaining the recommendations to the users is believed to improve transparency and to instill trust in the users. So far we have pursued our main goal in improving the accuracy of the recommendations, and in this paper we showed how we could achieve significant improvements by taking the temporal aspects into consideration. As the next step, in particular we are interested in how we can explain the temporal properties of the trained models to the users. Furthermore, the component-based structure followed in designing Aspect-MF is generally beneficial in extracting explanations.

\begin{acks}
	\label{Acknowledgment}
	We would like to acknowledge the SunCorp Group for partially funding this project. We would also like to thank the National eResearch Collaboration Tools and Resources (Nectar) for providing us with the necessary computational resources to carry out the experiments.
\end{acks}

\bibliographystyle{ACM-Reference-Format}
\bibliography{references}

\appendix
\section{Aspect-MF training equations}
\label{Aspect-MF Training Equations}

In Aspect-MF, we use gradient descent to optimise Eq. \ref{eq19}. The gradients for the model parameters are obtained using Eqs. \ref{eqa20} to \ref{eqa41}.

\small
\begin{equation}
	\begin{split}
		\label{eqa1}
		\frac{\partial E}{\partial bu_{u}} = \frac{\partial E_R}{\partial bu_{u}} = e_{uj} + \lambda_{bu}|I_u|^{-\frac{1}{2}}bu_{u} 
	\end{split}
\end{equation}
\normalsize

\small
\begin{equation}
	\begin{split}
		\label{eqa2}
		\forall {t_{uj}}\in I_u^t:
		\frac{\partial E}{\partial but_{ut}} = \frac{\partial E_R}{\partial but_{ut}} = e_{uj} + \lambda_{bu}|I_u|^{-\frac{1}{2}}but_{ut} 
	\end{split}
\end{equation}
\normalsize

\small
\begin{equation}
	\begin{split}
		\label{eqa3}
		\frac{\partial E}{\partial \alpha_{u}} = \frac{\partial E_R}{\partial \alpha_{u}} = e_{uj} dev_u(t_{uj}) + \lambda_{bu}|I_u|^{-\frac{1}{2}}\alpha_{u}
	\end{split}
\end{equation}
\normalsize

\small
\begin{equation}
	\begin{split}
		\label{eqa4}
		\frac{\partial E}{\partial bi_{j}} = \frac{\partial E_R}{\partial bi_{j}} = e_{uj} (C_u + C_{ut}) + \lambda_{bi}|J_j|^{-\frac{1}{2}}bi_{j}
	\end{split}
\end{equation}
\normalsize

\small
\begin{equation}
	\begin{split}
		\label{eqa5}
		\frac{\partial E}{\partial bit_{jBin(t_{uj})}} = \frac{\partial E_R}{\partial bit_{jBin(t_{uj})}} = e_{uj} (C_u + C_{ut}) + \lambda_{bi}|J_j|^{-\frac{1}{2}}bit_{jBin(t_{uj})}
	\end{split}
\end{equation}
\normalsize

\small
\begin{equation}
	\label{eqa6}
	\begin{split}
		\frac{\partial E}{\partial C_{u}} = \frac{\partial E_R}{\partial C_{u}} = e_{uj} (bi_j + bit_{jBin(t_{uj})}) + \lambda_{bi}|J_j|^{-\frac{1}{2}}C_{u}
	\end{split}
\end{equation}
\normalsize

\small
\begin{equation}
	\begin{split}
		\label{eqa7}
		\frac{\partial E}{\partial Ct_{u}} = \frac{\partial E_R}{\partial Ct_{u}} = e_{uj} (bi_j + bit_{jBin(t_{uj})}) + \lambda_{bi}|J_j|^{-\frac{1}{2}}Ct_{u}
	\end{split}
\end{equation}
\normalsize

\small
\begin{equation}
	\begin{split}
		\label{eqa8}
		\frac{\partial E}{\partial P_{uf}}=\frac{\partial E_R}{\partial P_{uf}}+\frac{\partial E_T}{\partial P_{uf}}
	\end{split} 
\end{equation}
\normalsize

\small
\begin{equation}
	\begin{split}
		\label{eqa9}
		\frac{\partial E_R}{\partial P_{uf}}=e_{uj}(W_{uf}(t_uj)Q_{jf}(t_uj) + Z_{uf}(t_uj)) + \lambda_P|I_u|^{-\frac{1}{2}}P_{uf}
	\end{split} 
\end{equation}
\normalsize

\small
\begin{equation}
	\begin{split}
		\label{eqa10}
		\frac{\partial E_T}{\partial P_{uf}}=\lambda_{T}|T_u|^{-\frac{1}{2}}P_{uf}+\lambda_t\eta_P\sum_{\forall v\in T_u}e^{(1)}_{uv}\omega_{vf}
	\end{split} 
\end{equation}
\normalsize

\small
\begin{equation}
	\begin{split}
		\label{eqa11}
		\forall {t_{uj}}\in I_u^t:
		\frac{\partial E}{\partial Pt_{uft}}=\frac{\partial E_R}{\partial Pt_{uft}}+\frac{\partial E_T}{\partial Pt_{uft}}
	\end{split}
\end{equation}
\normalsize

\small
\begin{equation}
	\begin{split}
		\label{eqa12}
		\frac{\partial E_R}{\partial Pt_{uft}}=e_{uj}(W_{uf}(t_uj)Q_{jf}(t_{uj})+Z_{uf}(t_uj))+\lambda_{Pt}|I_u|^{-\frac{1}{2}}Pt_{uft}
	\end{split}
\end{equation}
\normalsize

\small
\begin{equation}
	\begin{split}
		\label{eqa13}
		\frac{\partial E_T}{\partial Pt_{uft}}=\lambda_{T}|T_u|^{-\frac{1}{2}}Pt_{uft} + \frac{\lambda_t \eta_P}{|I_u^t|}\sum_{\forall v\in T_u}e^{(1)}_{uv}\omega_{vf} 
	\end{split}
\end{equation}
\normalsize

\small
\begin{equation}
	\label{eqa14}
	\begin{split}
		\frac{\partial E}{\partial \alpha_{uf}^P}=\frac{\partial E_R}{\partial \alpha_{uf}^P}+\frac{\partial E_T}{\partial \alpha_{uf}^P}
	\end{split}
\end{equation}
\normalsize

\small
\begin{equation}
	\label{eqa15}
	\begin{split}
		\frac{\partial E_R}{\partial \alpha_{uf}^P}=e_{uj}dev_u(t_{uj})(W_{uf}(t_uj)Q_{jf}(t_uj) + Z_{uf}(t_uj))+\lambda_{\alpha^P}|I_u|^{-\frac{1}{2}}\alpha_{uf}^P
	\end{split}
\end{equation}
\normalsize

\small
\begin{equation}
	\label{eqa16}
	\begin{split}
		\frac{\partial E_T}{\partial \alpha_{uf}^P}=\lambda_{T}|T_u|^{-\frac{1}{2}}\alpha_{uf}^P+\frac{\lambda_t \eta_P}{|I_u^t|}\sum_{\forall v\in T_u}\sum_{{\forall t_{uj}} \in I_u^t}e^{(1)}_{uv}\omega_{vf}dev_u(t_{uj})
	\end{split}
\end{equation}
\normalsize

\small
\begin{equation}
	\begin{split}
		\label{eqa17}
		\frac{\partial E}{\partial W_{uf}}=\frac{\partial E_R}{\partial W_{uf}}+\frac{\partial E_T}{\partial W_{uf}}
	\end{split}
\end{equation}
\normalsize

\small
\begin{equation}
	\begin{split}
		\label{eqa18}
		\frac{\partial E_R}{\partial W_{uf}}&=e_{uj}Q_{jf}(t_uj)(W_{uf}(t_uj)Q_{jf}(t_uj)+Z_{uf}(t_uj))\\&+2Q_{jf}(t_uj)\sum_{f^{'}=1}^{D}(W_{uf}(t_uj)Q_{jf^{'}}(t_uj) + Z_{uf^{'}}(t_uj))+\lambda_W|I_u|^{-\frac{1}{2}}W_{uf}
	\end{split}
\end{equation}
\normalsize

\small
\begin{equation}
	\begin{split}
		\label{eqa19}
		\frac{\partial E_T}{\partial W_{uf}}=\lambda_{T}|T_u|^{-\frac{1}{2}}W_{uf}+\lambda_t \eta_W\sum_{\forall v\in T_u}e^{(2)}_{uv}\omega_{vf} 
	\end{split}
\end{equation}
\normalsize

\small
\begin{equation}
	\begin{split}
		\label{eqa20}
		\forall {t_{uj}}\in I_u^t:
		\frac{\partial E}{\partial Wt_{uft}}=\frac{\partial E_R}{\partial Wt_{uft}}+\frac{\partial E_T}{\partial Wt_{uft}}
	\end{split}
\end{equation}
\normalsize

\small
\begin{equation}
	\begin{split}
		\label{eqa21}
		\frac{\partial E_R}{\partial Wt_{uft}}&=e_{uj}Q_{jf}(t_uj)(W_{uf}(t_uj)Q_{jf}(t_{uj}) + Z_{uf}(t_uj))\\&+2Q_{jf}(t_uj)\sum_{f^{'}=1}^{D}(W_{uf}(t_uj)Q_{jf^{'}}(t_uj) + Z_{uf^{'}}(t_uj))+\lambda_{Wt}|I_u|^{-\frac{1}{2}}W_{uf}
	\end{split}
\end{equation}
\normalsize

\small
\begin{equation}
	\begin{split}
		\label{eqa22}
		\frac{\partial E_T}{\partial Wt_{uft}}=\lambda_{T}|T_u|^{-\frac{1}{2}}W_{uf}+\frac{\lambda_t \eta_W}{|I_u^t|}\sum_{\forall v\in T_u}e^{(2)}_{uv}\omega_{vf} 
	\end{split}
\end{equation}
\normalsize

\small
\begin{equation}
	\begin{split}
		\label{eqa23}
		\frac{\partial E}{\partial \alpha_{uf}^W}=\frac{\partial E_R}{\partial \alpha_{uf}^W}+\frac{\partial E_T}{\partial \alpha_{uf}^W}
	\end{split}
\end{equation}
\normalsize

\small
\begin{equation}
	\begin{split}
		\label{eqa24}
		\frac{\partial E_R}{\partial \alpha_{uf}^W}&=e_{uj}dev_u(t_{uj})(W_{uf}(t_uj)Q_{jf}(t_uj) + Z_{uf}(t_uj))\\&+2Q_{jf}(t_uj)\sum_{f^{'}=1}^{D}(W_{uf}(t_uj)Q_{jf^{'}}(t_uj)+Z_{uf^{'}}(t_uj))+\lambda_{\alpha^W}|I_u|^{-\frac{1}{2}}\alpha_{uf}^W
	\end{split}
\end{equation}
\normalsize

\small
\begin{equation}
	\begin{split}
		\label{eqa25}
		\frac{\partial E_T}{\partial \alpha_{uf}^W}=\lambda_{T}|T_u|^{-\frac{1}{2}}\alpha_{uf}^W+\frac{\lambda_t \eta_W}{|I_u^t|}\sum_{\forall v\in T_u}\sum_{\forall {t_{uj}} \in I_u^t}e^{(2)}_{uv}\omega_{vf}dev_u(t_{uj})
	\end{split}
\end{equation}
\normalsize

\small
\begin{equation}
	\begin{split}
		\label{eqa26}
		\frac{\partial E}{\partial Z_{uf}}=\frac{\partial E_R}{\partial Z_{uf}}+\frac{\partial E_T}{\partial Z_{uf}}
	\end{split}
\end{equation}
\normalsize

\small
\begin{equation}
	\begin{split}
		\label{eqa27}
		\frac{\partial E_R}{\partial Z_{uf}}&=e_{uj}(W_{uf}(t_uj)Q_{jf}(t_uj) + Z_{uf}(t_uj))+2\sum_{f^{'}=1}^{D}(W_{uf}(t_uj)Q_{jf^{'}}(t_uj)+Z_{uf^{'}}(t_uj))\\&+\lambda_Z|I_u|^{-\frac{1}{2}}Z_{uf}
	\end{split}
\end{equation}
\normalsize

\small
\begin{equation}
	\begin{split}
		\label{eqa28}
		\frac{\partial E_T}{\partial Z_{uf}}=\lambda_{T}|T_u|^{-\frac{1}{2}}Z_{uf}+\lambda_t\eta_Z\sum_{\forall v\in T_u}e^{(3)}_{uv}\omega_{vf} 
	\end{split}
\end{equation}
\normalsize

\small
\begin{equation}
	\begin{split}
		\label{eqa29}
		\forall {t_{uj}}\in I_u^t:
		\frac{\partial E}{\partial Zt_{uft}}=\frac{\partial E_R}{\partial Zt_{uft}}+\frac{\partial E_T}{\partial Zt_{uft}}
	\end{split}
\end{equation}
\normalsize

\small
\begin{equation}
	\begin{split}
		\label{eqa30}
		\frac{\partial E_R}{\partial Zt_{uft}}=&e_{uj}(W_{uf}(t_uj)Q_{jf}(t_{uj}) + Z_{uf}(t_uj)) + 2\sum_{f^{'}=1}^{D}(W_{uf}(t_uj)Q_{jf^{'}}(t_uj) + Z_{uf^{'}}(t_uj))+\\&\lambda_Z|I_u|^{-\frac{1}{2}}Zt_{uft}
	\end{split}
\end{equation}
\normalsize

\small
\begin{equation}
	\begin{split}
		\label{eqa31}
		\frac{\partial E_T}{\partial Zt_{uft}}=\lambda_{T}|T_u|^{-\frac{1}{2}}Zt_{uft}+\frac{\lambda_t \eta_Z}{|I_u^t|}\sum_{\forall v\in T_u}e^{(3)}_{uv}\omega_{vf} 
	\end{split}
\end{equation}
\normalsize

\small
\begin{equation}
	\begin{split}
		\label{eqa32}
		\frac{\partial E}{\partial \alpha_{uf}^Z}=\frac{\partial E_R}{\partial \alpha_{uf}^Z}+\frac{\partial E_T}{\partial \alpha_{uf}^Z}
	\end{split}
\end{equation}
\normalsize

\small
\begin{equation}
	\begin{split}
		\label{eqa33}
		\frac{\partial E_R}{\partial \alpha_{uf}^Z}&=e_{uj}dev_u(t_{uj})(W_{uf}(t_uj)Q_{jf}(t_uj) + Z_{uf}(t_uj))\\&+2\sum_{f^{'}=1}^{D}(W_{uf}(t_uj)Q_{jf^{'}}(t_uj) + Z_{uf^{'}}(t_uj))+\lambda_{\alpha^Z}|I_u|^{-\frac{1}{2}}\alpha_{uf}^Z
	\end{split}
\end{equation}
\normalsize

\small
\begin{equation}
	\begin{split}
		\label{eqa34}
		\frac{\partial E_T}{\partial \alpha_{uf}^Z}=\lambda_{T}|T_u|^{-\frac{1}{2}}\alpha_{uf}^Z+\frac{\lambda_t \eta_Z}{|I_u^t|}\sum_{\forall v\in T_u}\sum_{\forall {t_{uj}} \in I_u^t}e^{(3)}_{uv}\omega_{vf}dev_u(t_{uj})
	\end{split}
\end{equation}
\normalsize

\small
\begin{equation}
	\label{eqa35}
	\begin{split}
		\forall i \in I_{u}:
		\frac{\partial E}{\partial y_{if}}=\frac{\partial E_R}{\partial y_{if}}=e_{uj}|I_u|^{-\frac{1}{2}}(W_{uf}V_{jf} + Z_{uf})+(\lambda_y|J_j|^{-\frac{1}{2}}y_{if})
	\end{split}
\end{equation}
\normalsize

\small
\begin{equation}
	\label{eqa36}
	\begin{split}
		\forall v \in T_{u}:
		\frac{\partial E}{\partial \omega_{vf}}=\frac{\partial E_R}{\partial \omega_{vf}}+\frac{\partial E_T}{\partial \omega_{vf}}
	\end{split}
\end{equation}
\normalsize

\small
\begin{equation}
	\label{eqa37}
	\begin{split}
		\frac{\partial E_R}{\partial \omega_{vf}}=e_{uj}|T_u|^{-\frac{1}{2}}(W(t)_{uf}Q_{jf} + Z(t)_{uf})\\&
	\end{split}
\end{equation}
\normalsize

\small
\begin{equation}
	\label{eqa38}
	\begin{split}
		\frac{\partial E_T}{\partial \omega_{vf}}&=(\lambda_{T}|T^{+}_{v}|^{-\frac{1}{2}})\omega_{vf}+\frac{\lambda_t \eta_P}{|I_u^t|}\sum_{\forall {t_{uj}} \in I_u^t}e^{(1)}_{uv}P(t_{uj})_{uf}\\&+\frac{\lambda_t \eta_W}{|I_u^t|}\sum_{\forall {t_{uj}} \in I_u^t}e^{(2)}_{uv}(1 - W(t_{uj})_{uf}) + \frac{\lambda_t \eta_Z}{|I_u^t|}\sum_{\forall {t_{uj}} \in I_u^t}e^{(3)}_{uv}Z(t_{uj})_{uf}
	\end{split}
\end{equation}
\normalsize

\small
\begin{equation}
	\begin{split}
		\label{eqa39}
		\frac{\partial E}{\partial Y_{ff^{'}}}=\frac{\partial E_R}{\partial Y_{ff^{'}}}=e_{uj}(W_{if}V_{jf} + Z_{if})(W_{if^{'}}V_{jf^{'}} + Z_{if^{'}}) - \lambda_{Y}Y_{ff^{'}}
	\end{split}
\end{equation}
\normalsize

\small
\begin{equation}
	\label{eqa40}
	\begin{split}
		\frac{\partial E}{\partial Q_{jf}}&=\frac{\partial E_R}{\partial Q_{jf}}=e_{uj}[W_{uf}(P_{uf} + |I_u|^{-\frac{1}{2}}\sum_{\forall i \in I_u}y_{i}+\\&|T_u|^{-\frac{1}{2}}\sum_{\forall v \in T_u}\omega_{v})+2W_{uf}\sum_{f^{'}=1}^{D}(W_{if^{'}}V_{jf^{'}} + Z_{if^{'}})Y_{ff^{'}}] + \lambda_Q|U_j|^{-\frac{1}{2}}Q_{jf} 
	\end{split}
\end{equation}
\normalsize

where:
\begin{equation}
	\label{eqa41}
	\begin{split}
		e_{uj}=R_{uj} - \hat{R_{uj}}
	\end{split}
\end{equation}
\normalsize

Therefore, the gradients in Eqs. \ref{eqa20} to \ref{eqa41} will be used to update the values matrices used to capture socially-influenced drifting conditional feature value preferences using an incremental gradient descent method.

\end{document}